\documentclass[useAMS, usenatbib, usegraphicx, a4paper]{mn2e}
\usepackage{amssymb}

\newcommand{\mb}[1]{\mbox{\boldmath $#1$}}
\newcommand{\q}{\quad}
\newcommand{\qq}{\qquad}

\title[Pulsations of rotating relativistic stars]
      {Non-linear axisymmetric pulsations of rotating relativistic
       stars in the conformal flatness approximation}

\author[H.~Dimmelmeier, N.~Stergioulas and J.~A.~Font]
       {Harald Dimmelmeier$^1$,
        Nikolaos Stergioulas$^2$\thanks{E-mail:niksterg@astro.auth.gr}
        and Jos\'e A.~Font$^3$ \\
        $^1$Max-Planck-Institut f\"ur Astrophysik,
            Karl-Schwarzschild-Strasse 1, D-85741 Garching, Germany \\
        $^2$Department of Physics, Aristotle University of
            Thessaloniki, Thessaloniki 54124, Greece \\
        $^3$Departamento de Astronom\'{\i}a y Astrof\'{\i}sica,
            Universidad de Valencia, Dr.~Moliner 50,
            46100 Burjassot (Valencia), Spain}

\begin{document}

\date{Accepted $<$date$>$. Received $<$date$>$; in original form $<$date$>$}

\pagerange{\pageref{firstpage}--\pageref{lastpage}} \pubyear{2005}

\label{firstpage}

\maketitle


\begin{abstract}
  We study non-linear axisymmetric pulsations of rotating relativistic
  stars using a general relativistic hydrodynamics code under the
  assumption of a conformally flat three-metric. We compare the
  results of our simulations, in which the spacetime dynamics is
  coupled to the evolution of the fluid, to previous results performed
  in the Cowling approximation in which the spacetime dynamics was
  neglected. We show that the conformal flatness condition has only a
  small effect on the dynamics of pulsating relativistic stars and the
  obtained pulsation frequencies are very close to those expected in
  full general relativity. The pulsations are studied along various
  sequences of both uniformly and differentially rotating relativistic
  polytropes with index $ N = 1 $. For small pulsation amplitudes we
  identify several modes, including the lowest-order $ l = 0 $, 2, and
  4 axisymmetric modes, as well as several axisymmetric inertial
  modes. Differential rotation significantly shifts mode frequencies
  to smaller values, increasing the likelihood of detection by current
  gravitational wave interferometric detectors. We observe an extended
  avoided crossing between the $ l = 0 $ and $ l = 4 $ first overtones
  (previously known to exist from perturbative studies), which is
  important for correctly identifying mode frequencies in case of
  detection. For uniformly rotating stars near the mass-shedding
  limit, we confirm the existence of the mass-shedding-induced damping
  of pulsations and argue that it is still relevant for secularly
  unstable modes, even though the effect is not as strong as was
  previously found in the Cowling approximation. We also investigate
  non-linear harmonics of the linear modes and notice that rotation
  changes the pulsation frequencies in a way that would allow for
  various parametric instabilities between two or three modes to take
  place. Although this scenario has been explored before for
  slowly-rotating collapse, it could become very interesting in the
  case of rapidly rotating collapse, where the quasi-radial mode could
  be in resonance with inertial modes. We assess the detectability of
  each obtained mode by current gravitational wave detectors and
  outline how the empirical relations that have been constructed for
  gravitational wave asteroseismology could be extended to include the
  effects of rotation.
\end{abstract}

\begin{keywords}
  Hydrodynamics -- relativity -- methods: numerical -- stars:
  neutron -- stars: oscillations -- stars: rotation
\end{keywords}


\section{Introduction}
\label{section:introduction}

Axisymmetric pulsations of rotating neutron stars could be excited in
a number of astrophysical scenarios, such as rotating core collapse,
accretion-induced collapse, core quakes due to a large
phase-transition in the equation of state (EOS), or hypermassive
neutron star formation in a binary neutron star merger \citep[see][for
extensive discussions] {stergioulas_04_a, kokkotas_05_a}. These
pulsations are a potential source of detectable high-frequency
gravitational waves. While for nonrotating stars the frequencies of
normal modes can be computed with perturbative methods and a theory of
gravitational wave asteroseismology has already been formulated
\citep{andersson_98_a, kokkotas_01_a, benhar_04_a}, there exist no
accurate frequency determinations for rapidly rotating stars to date,
nor has the theory of gravitational wave asteroseismology been
extended to include the effects of rotation on the oscillation
frequencies. Most existing computations of oscillation modes in
rapidly rotating relativistic stars use the Cowling approximation
\citep[]{yoshida_02_a, stergioulas_04_a, yoshida_05_a}, with the only
exception being the computation of the two lowest-order quasi-radial
modes in \citet{font_02_a} \citep[see also][]{shibata_03_a,
  shibata_03_b}.

In this paper we present an accurate determination of several
axisymmetric pulsation modes of rotating stars in general
relativity. The accurate knowledge of the frequencies of different
modes excited in the astrophysical events mentioned above is necessary
both for narrow-banding of the detectors, as well as for solving the
inverse problem, i.e.\ identifying the EOS of high-density matter.

The traditional approach for computing mode frequencies uses
perturbation theory for either solving a time-independent eigenvalue
problem or for obtaining the time evolution of the linearised
equations governing the dynamics of matter and spacetime
\citep[see][for comprehensive reviews of these
approaches]{kokkotas_99_a, kokkotas_03_a}. The advantage of the
perturbative approach is that the equations can be expanded in terms
of spherical harmonics. However, for rapidly rotating relativistic
stars this approach has only worked in the Cowling approximation so
far \citep{yoshida_99_a, yoshida_01_a, yoshida_02_a, yoshida_05_a},
except for zero-frequency $ f $-modes, which were computed in full
general relativity \citep{stergioulas_98_a, morsink_99_a}. The main
problem for applying the perturbative approach in full general
relativity is the absence of analytic boundary conditions at infinity,
which would allow to apply the outgoing-wave boundary conditions
defining the quasi-normal modes \citep[see][for a
review]{stergioulas_03_a}. Only if one assumes the slow-rotation
approximation the problem is still tractable \citep[see, e.g.,][]
{hartle_75_a, kojima_97_a, datta_98_a, ruoff_02_a, pons_05_a}.

In recent years, the time evolution of the non-linear equations
governing the dynamics of matter and spacetime has been introduced
as a promising new approach for computing mode frequencies
\citep{font_00_a, font_01_a, font_02_a, stergioulas_01_a,
  stergioulas_04_a}. For small amplitudes, the obtained frequencies
are in excellent agreement with those expected by linear perturbation
theory, while two-dimensional eigenfunctions can be obtained through a
Fourier transform technique \citep[see][SAF
hereafter]{stergioulas_04_a}. The advantages of this method are that
one does not need precise outer boundary conditions and that one can
also study non-linear pulsations.

This study extends the results presented in SAF (which were obtained
in the Cowling approximation) by incorporating the spacetime dynamics
in the evolutions. This is done by using the Isenberg--Wilson--Mathews
approximation of general relativity (also known as the \emph{conformal
  flatness condition}; hereafter CFC) where the 3\,+\,1 Einstein
equations reduce to a non-linear set of five coupled elliptic
equations for the lapse function, the shift vector, and the conformal
factor \citep{isenberg_78_a, wilson_96_a}. The approximation
essentially ignores gravitational radiation and is thus appropriate
for equilibrium, quasi-equilibrium, but also for highly dynamical
situations \citep[see, e.g.,][]{cook_96_a, dimmelmeier_02_a,
  oechslin_02_a, faber_04_a, saijo_04_a}.

For small pulsation amplitudes we identify several modes, including
the lowest-order $ l = 0 $, 2, and 4 axisymmetric modes as well as
several axisymmetric inertial modes. The pulsations are studied along
the same sequences of uniformly and differentially rotating
relativistic polytropes with index $ N = 1 $ as in SAF. Differential
rotation significantly shifts mode frequencies to smaller values,
increasing the likelihood of detection by current gravitational wave
interferometric detectors. An important feature of the frequency
spectrum, induced by rotation, is the existence of avoided crossings
between different mode sequences \citep[see][]{clement_86_a,
  yoshida_01_a}. We observe an extended avoided crossing between the
$ l = 0 $ and $ l = 4 $ first overtones. This is important for
correctly identifying mode frequencies in case of detection.

Our non-linear approach allows us to identify non-linear harmonics
in addition to the well known linear modes. These harmonics arise
due to couplings between various modes or due to non-linear
self-couplings \citep[see also][]{sperhake_01_a, sperhake_02_a};
for similar results obtained for the non-linear oscillations of a torus
orbiting a black hole, see \citet{zanotti_05_a}. It has been suggested
\citep{clark_79_a, eardley_83_a} that nonradial oscillations after
core bounce could be enhanced through a parametric instability with
the quasi-radial mode \citep[see also][for recent related
work]{passamonti_05_a}. In nonrotating or slowly rotating collapse,
such a parametric instability can only take place under special
conditions that would allow the two modes to be in resonance. In our
work we find that rotational shifting of the frequency of different
modes broadens the range of parameters for which interesting
resonances could take place. In particular we notice that the
quasi-radial mode will be in resonance with some inertial mode(s) for
all rotation rates above a critical value. It is thus interesting to
further study the possible energy transfer between different modes
excited after, e.g., a core collapse or an accretion-induced collapse
event, either on secular time-scales or as a possible parametric
instability.

The paper is organized as follows. In Section~\ref{section:framework}
we introduce the mathematical and numerical framework and specify the
initial fluid perturbations, while in
Section~\ref{section:equilibrium_models} we present the sequences of
equilibrium initial models. In Section~\ref{section:linear_pulsations}
we discuss the effects of linear pulsations, focusing on the role of
rotation and avoided crossings of modes, and present a detailed
comparison to results in the Cowling
approximation. Section~\ref{section:recycling} is devoted to the
technique of mode recycling, and in
Section~\ref{section:nonlinear_pulsations} we examine non-linear
effects of the pulsations like mode coupling or mass-shedding-induced
damping. Gravitational wave emission and asteroseismology are
discussed in Section~\ref{section:gravitational_waves}. A summary of
our results in Section~\ref{section:summary} concludes this work.

Unless otherwise noted, we choose dimensionless units for all physical
quantities by setting the speed of light, the gravitational constant,
and the solar mass to one, $ c = G = M_\odot = 1 $. Latin indices run
from 1 to 3, Greek indices from 1 to 4.


\section{Mathematical and numerical framework}
\label{section:framework}

We study axisymmetric pulsations of rapidly rotating relativistic
stars by first constructing several sequences of uniformly and
differentially rotating equilibrium models. In SAF the equilibrium
models were constructed using the numerical code
\texttt{rns}~\citep{stergioulas_95_a}. In the present work we build
the stellar equilibrium models using the self-consistent field method
described in \citet{komatsu_89_a, komatsu_89_b} (KEH hereafter), which
solves the general relativistic hydrostatic equations for rotating
matter distributions whose pressure obeys an EOS given by a polytropic
relation (see Eq.~(\ref{eq:polytropic_eos}) below). Comparisons of the
accuracy of both approaches in the case of uniform rotation can be
found in \citet{nozawa_98_a} and in~\citet{stergioulas_03_a}. Specific
details of the equilibrium models are deferred to
Section~\ref{section:equilibrium_models} below, where a quantitative
comparison of the equilibrium properties of a particular highly
differentially rotating model built either using \texttt{rns} or the
KEH solver is made.

These equilibrium models are taken as initial data for the evolution
code after a suitable perturbation has been added in order to excite
specific modes of oscillation (see
Section~\ref{subsection:perturbations}).
The time dependent numerical simulations are performed with the code
\textsc{CoCoNuT}, developed by \citet{dimmelmeier_02_a,
dimmelmeier_02_b} with a metric solver based on spectral methods as
described in \citet{dimmelmeier_05_a}. The code uses the general
relativistic field equations for a curved spacetime in the
3\,+\,1-split under the assumption of conformal flatness for the
three-metric. The hydrodynamics equations are consistently formulated
in conservation form, and are solved by high-resolution
shock-capturing schemes.

In the code used by SAF, the spacetime dynamics was neglected, and
the attention was focused on the oscillations of the stars on a fixed
background metric given by the solution of the Einstein equations on
the initial time slice. Keeping the spacetime fixed to the initial
equilibrium state during the evolution corresponds to the Cowling
approximation in perturbation theory. In the simulations presented in
this work, however, such a simplification is not made and the
spacetime fields are also allowed to evolve in time.

In the following, we present the mathematical formulation of the
metric and hydrodynamics equations, and then summarise the numerical
methods used for solving them.


\subsection{Metric equations}
\label{subsection:metric_equations}

We adopt the ADM 3\,+\,1 formalism by \citet{arnowitt_62_a} to
foliate a spacetime endowed with a metric $ g_{\mu\nu} $ into a set of
non-intersecting spacelike hypersurfaces. The line element then reads
\begin{equation}
  ds^2 = g_{\mu\nu} \, dx^\mu \! dx^\nu = - \alpha^2 dt^2 +
  \gamma_{ij} (dx^i + \beta^i dt) (dx^j + \beta^j dt), ~
  \label{eq:line_element}
\end{equation}
where $ \alpha $ is the lapse function, $ \beta^i $ is the spacelike
shift three-vector, and $ \gamma_{ij} $ is the spatial three-metric.

In the 3\,+\,1 formalism, the Einstein equations are split into
evolution equations for the three-metric $ \gamma_{ij} $ and the
extrinsic curvature $ K_{ij} $, and constraint equations (the
Hamiltonian and momentum constraints) which must be fulfilled at every
spacelike hypersurface:
\begin{equation}
  \setlength{\arraycolsep}{0.14 em}
  \begin{array}{rcl}
    \partial_t \gamma_{ij} & = & - 2 \alpha K_{ij} +
    \nabla_{\!i} \beta_j + \nabla_{\!j} \beta_i,
    \\ [0.8 em]
    \partial_t K_{ij} & = & - \nabla_{\!i} \nabla_{\!j} \alpha +
    \alpha (R_{ij} \!-\! 2 K_{ik} K_j^k) +
    \beta^k \nabla_{\!k} K_{ij} + K_{ik} \! \nabla_{\!j} \beta^k
    \hspace{-2 em}
    \\ [0.5 em]
    & & \displaystyle +
    K_{jk} \nabla_{\!i} \beta^k - 8 \pi \alpha \! \left( \! S_{ij} -
    \frac{\gamma_{ij}}{2} (S_k^k - \rho_\mathrm{H}) \right),
    \\ [0.8 em]
    0 & = & R - K_{ij} K^{ij} - 16 \pi \rho_\mathrm{H},
    \\ [0.8 em]
    0 & = & \nabla_{\!i} K^{ij} - 8 \pi S^j,
  \end{array}
  \label{eq:adm_metric_equations}
\end{equation}
where the maximal slicing condition, $ K^i_i = 0 $, is imposed. In
these equations $ \nabla_{\!i} $ is the covariant derivative with
respect to the three-metric $ \gamma_{ij} $, $ R_{ij} $ is the
corresponding Ricci tensor, and $ R $ is the scalar curvature.

The matter fields of the general relativistic fluid appearing in the
above equations are the spatial components $ S_{ij} $ of the (perfect
fluid) stress-energy tensor $ T^{\mu\nu} = \rho h u^\mu u^\nu + P 
g^{\mu\nu}$, the three momenta $ S^i = \rho h W^2 v^i $, and the total
energy $ \rho_\mathrm{H} = \alpha^2 T^{00} $. The fluid is specified
by the rest-mass density $ \rho $, the four-velocity $ u^\mu $, and
the pressure $ P $, with the specific enthalpy defined as
$ h = 1 + \epsilon + P / \rho $, where $ \epsilon $ is the specific
internal energy. The three-velocity of the fluid as measured by an
Eulerian observer is given by
$ v^i = u^i / (\alpha u^0) + \beta^i / \alpha $, and the Lorentz
factor $ W = \alpha u^0 $ satisfies the relation
$ W = 1 / \sqrt{1 - v_i v^i} $.

The equations of the original ADM formulation, when implemented 
numerically with standard finite-difference methods, suffer from
several numerical instabilities. For many years there have been
numerous attempts to reformulate these equations into forms better
suited for numerical work \citep[see, e.g.,][and references
therein]{alcubierre_04_a}. One recent approach is based on a
constrained evolution scheme \citep{bonazzola_04_a}, which exploits
the observation that the more constraints are used in the formulation
of the equations the more numerically stable the evolution appears to
be. We refer the interested reader to the discussion in
\citet{dimmelmeier_05_a} and references therein.

Based on the ideas of \citet{isenberg_78_a} and \citet{wilson_96_a},
and as it was done in the work of
\citet{dimmelmeier_02_a, dimmelmeier_02_b}, we follow a similar
strategy. We approximate the general metric $ g_{\mu\nu} $
by replacing the spatial three-metric $ \gamma_{ij} $ with the
conformally flat three-metric
\begin{equation}
  \gamma_{ij} = \phi^4 \hat{\gamma}_{ij},
  \label{eq:cfc_metric}
\end{equation}
where $ \hat{\gamma}_{ij} $ is the flat metric and $\phi$ is the
conformal factor. Therefore, at all times during a numerical
simulation we assume that all off-diagonal components of the
three-metric are zero, and the diagonal elements have the common
factor $ \phi^4 $.

In this CFC approximation the expression for the extrinsic curvature
becomes time-independent and reads
\begin{equation}
  K_{ij} =
  \frac {1}{2 \alpha} \left( \nabla_{\!i} \beta_j + \nabla_{\!j}
  \beta_i - \frac{2}{3} \gamma_{ij} \nabla_{\!k} \beta^k \right).
  \label{eq:extrinsic_curvature_cfc}
\end{equation}
With this the ADM equations~(\ref{eq:adm_metric_equations}) reduce to
a set of five coupled elliptic non-linear equations for the metric
components,
\begin{equation}
  \setlength{\arraycolsep}{0.14 em}
  \begin{array}{rcl}
    \hat{\Delta} \phi & = & \displaystyle - 2 \pi \phi^5 \left( \rho h
    W^2 - P + \frac{K_{ij} K^{ij}}{16 \pi} \right), \\ [1.0 em]
    \hat{\Delta} (\alpha \phi) & = & \displaystyle 2 \pi \alpha \phi^5
    \! \left( \! \rho h (3 W^2 - 2) \! + \! 5 P \! + \! \frac{7 K_{ij}
    K^{ij}}{16 \pi} \! \right), \!\!\!\!\!\!\!\!\!\!\!\! \\ [1.0 em]
    \hat{\Delta} \beta^i & = & \displaystyle 16 \pi \alpha \phi^4 S^i 
+
    2 \phi^{10} K^{ij} \hat{\nabla}_{\!j}
    \left( \frac{\alpha}{\phi^6} \right) -
    \frac{1}{3} \hat{\nabla}^i \hat{\nabla}_{\!k} \beta^k,
  \end{array}
  \label{eq:cfc_metric_equations}
\end{equation}
where $ \hat{\nabla}_{\!i} $ and $ \hat{\Delta} $ are the flat space
Nabla and Laplace operators, respectively. They do not contain
explicit time derivatives, and thus the metric is calculated by a
fully constrained approach, at the cost of neglecting some
evolutionary degrees of freedom in the spacetime metric (e.g.,
dynamical gravitational wave degrees of freedom). On each time slice
the metric is hence solely determined by the instantaneous
hydrodynamic state, i.e.\ the distribution of matter in space.

The accuracy of the CFC approximation has been tested in various
works, both in the context of stellar core collapse and for
equilibrium models of neutron stars~\citep{cook_96_a,
dimmelmeier_02_a, shibata_04_a, dimmelmeier_05_a, saijo_05_a,
  cerda_05_a}. The spacetime of rapidly (uniformly or differentially)
rotating neutron star models is still very well approximated by the
CFC metric~(\ref{eq:cfc_metric}). The accuracy of the approximation is
expected to degrade only in extreme cases, such as a rapidly rotating
black hole.

Recently, \citet{cerda_05_a} have extended the CFC system of equations
by incorporating additional degrees of freedom in the approximation,
which render the spacetime metric exact up to the second
post-Newtonian order (CFC+ approach). Results for uniformly rotating
pulsating neutron stars show only minute differences with respect to
the CFC approximation for the computed frequencies of the quasi-radial
fundamental $ F $-mode and its first two overtones. Moreover, a direct
comparison of the CFC approach with fully general relativistic
simulations of the quasi-radial modes of a rotating star, obtained
in \citet{font_02_a} have also yielded excellent agreement in the
oscillation frequencies.


\subsection{General relativistic hydrodynamics}
\label{subsection:gr_hydrodynamics}

The hydrodynamic evolution of a relativistic perfect fluid is
determined by a system of local conservation equations, which read
\begin{equation}
  \nabla_{\!\mu} J^{\mu} = 0, \qquad \nabla_{\!\mu} T^{\mu \nu} = 0,
  \label{eq:gr_equations_of_motion}
\end{equation}
where $ J^{\mu} = \rho u^{\mu} $ is the rest-mass current, and
$\nabla_{\!\mu}$ denotes the covariant derivative with respect to the
four-metric $ g_{\mu \nu} $. Following~\citet{banyuls_97_a} we 
introduce a set of conserved variables in terms of the primitive
(physical) variables $ (\rho, v_i, \epsilon) $:
\begin{displaymath}
  D = \rho W,
  \qquad
  S_i = \rho h W^2 v_i
  \qquad
  \tau = \rho h W^2 - P - D.
\end{displaymath}
Using the above variables, the local conservation
laws~(\ref{eq:gr_equations_of_motion}) can be written as a
first-order, flux-conservative hyperbolic system of equations,
\begin{equation}
  \frac{1}{\sqrt{- g}} \left[
  \frac{\partial \sqrt{\gamma} \mb{U}}{\partial t} +
  \frac{\partial \sqrt{- g} \mb{F}^i}{\partial x^i} \right] = \mb{S},
  \label{eq:hydro_conservation_equation}
\end{equation}
with the state vector, flux vector, and source vector
\begin{equation}
  \setlength{\arraycolsep}{0.14 em}
  \begin{array}{rcl}
  \mb{U} & = & [D, S_j, \tau], \\ [1.0 em]
  \mb{F}^i & = & \displaystyle
  \left[ D \hat{v}^i, S_j \hat{v}^i + \delta^i_j P,
  \tau \hat{v}^i + P v^i \right], \\ [1.0 em]
  \mb{S} & = & \displaystyle
  \left[ 0, T^{\mu \nu} \left(
  \frac{\partial g_{\nu j}}{\partial x^\mu} -
  {\it \Gamma}^\lambda_{\mu \nu} g_{\lambda j} \!\right),
  \alpha \left( T^{\mu 0}
  \frac{\partial \ln \alpha}{\partial x^\mu} -
  T^{\mu \nu} {\it \Gamma}^0_{\mu \nu} \!\right) \right]\!,
  \hspace{-2.5 em}
  \end{array}
  \label{eq:hydro_conservation_equation_constituents}
\end{equation}
respectively. Here $ \hat{v}^i = v^i - \beta^i / \alpha $, and
$ \sqrt{-g} = \alpha \sqrt{\gamma} $, with
$ g = \det (g_{\mu \nu}) $ and $ \gamma = \det (\gamma_{ij}) $. In
addition, $ {\it \Gamma}^\lambda_{\mu \nu} $ are the Christoffel
symbols associated with $ g_{\mu \nu} $.

The system of hydrodynamics
equations~(\ref{eq:hydro_conservation_equation}) is closed by an
EOS, which relates the pressure to some thermodynamically independent
quantities, e.g., $ P = P (\rho, \epsilon) $. For rotating neutron
star models below the mass-shedding limit \citep[see][for the precise
definition of the mass-shedding limit in the case of rapidly rotating
relativistic stars]{friedman_86_a} we assume that the star remains
isentropic. We can thus demand that the pressure obeys an EOS given by
the polytropic relation
\begin{equation}
  P = K \rho^\gamma,
  \label{eq:polytropic_eos}
\end{equation}
where $ K $ is the polytropic constant and $ \gamma = 1 + 1 / N $ is
the adiabatic index. Note that the evolution equation for the
generalised energy $ \tau $ can be discarded if an EOS of the form
$ P = P (\rho) $ as in Eq.~(\ref{eq:polytropic_eos}) is used. In this
particular case the internal specific energy can be obtained from the
ideal fluid EOS as $ \epsilon = P / [\rho (\gamma - 1)] $.

Near the mass-shedding limit, even small amplitude oscillations can
result in significant shedding of matter from the stellar surface in
the form of shocks (see Section~\ref{subsection:mass_shedding}). In
this case the polytropic relation~(\ref{eq:polytropic_eos}) does not
hold anymore. Therefore, as in the work by SAF, we then employ the 
adiabatic ideal fluid EOS instead,
\begin{equation}
  P = \rho \epsilon (\gamma - 1).
  \label{eq:ideal_gas_eos}
\end{equation}
We point out that at the initial time the isentropic equilibrium
models constructed with the polytropic EOS~(\ref{eq:polytropic_eos})
are consistent with the ideal fluid EOS~(\ref{eq:ideal_gas_eos}).


\subsection{Numerical methods for solving the metric and hydrodynamics
  equations}
\label{subsection:numerical_methods}

The hydrodynamics solver performs the numerical time integration of
the system of conservation
equations~(\ref{eq:hydro_conservation_equation}) using a
high-resolution shock-capturing (HRSC) scheme on a finite difference
grid \citep[for a review of such methods in numerical general
relativity, see][]{font_03_a}. This method ensures numerical
conservation of physically conserved quantities and a correct
treatment of discontinuities such as shocks (which may be present in
hydrodynamic quantities). In (upwind) HRSC methods a Riemann problem
has to be solved at each cell interface, which requires the
reconstruction of the primitive variables $ (\rho, v^i, \epsilon) $ at
these interfaces. We use the PPM method for the reconstruction, which
yields third order accuracy in space. The solution of the Riemann
problems then provides the numerical fluxes at cell interfaces. To
obtain this solution, the characteristic structure of the
hydrodynamics equations is explicitly needed \citep{banyuls_97_a}. In
our code the numerical fluxes are computed by means of Marquina's
approximate flux formula \citep{donat_98_a}. The time update of the
conserved vector $ \mb U $ is done using the method of lines in
combination with a Runge--Kutta scheme with second order accuracy in
time. Once the state vector is updated in time, the primitive
variables are recovered through an iterative Newton--Raphson method.

Although the most common approach to numerically solve the Einstein
equations is by means of finite differences, such methods are not
particularly well suited if the metric equations are formulated as
non-linear coupled elliptic equations like in the CFC approach. For
multidimensional simulations, the necessary grid resolutions typically
require to solve computationally very expensive numerical problems. If
iterative solvers in combination with spherical polar coordinates are
used, an additional obstacle may manifest itself in slow or failed
convergence due to problems at the coordinate origin or axis
\citep[see discussion in][]{dimmelmeier_05_a}. A possible remedy to
these shortcomings can be the use of non-linear multigrid solvers.

As an alternative strategy to reduce the complexity associated with
solving elliptic equations by reducing the number of grid points
required for a given numerical accuracy, we utilise an iterative
non-linear solver based on spectral methods in our code. This metric
solver, which is described in detail in \citet{dimmelmeier_05_a}, uses
routines from the publicly available object-oriented \textsc{Lorene}
library (\texttt{www.lorene.obspm.fr}), which supplies routines that
implement spectral methods in spherical polar coordinates. In contrast
to the hydrodynamic quantities, the metric components are always
smooth, and thus spectral methods are ideally suited for numerically
representing the spacetime metric. The practical implementation of the
combination of HRSC methods for the hydrodynamics and spectral methods
for the metric equations (the {\em Mariage des Maillages\/} or `grid
wedding' approach) in a multidimensional numerical code has been
presented in \citet{dimmelmeier_05_a}. 

The \textsc{CoCoNuT} code utilises Eulerian spherical polar
coordinates $ \{r, \theta, \varphi \} $, and thus axially or
spherically symmetric configurations can be easily simulated. For the
rotating neutron star models discussed in this work, we choose an
axisymmetric grid setup (no dependence of quantities on the coordinate
$ \varphi $), and assume symmetry with respect to the equatorial
plane. The finite difference grid consists of 160 radial and 60
angular grid points, which are equidistantly spaced. A small part of
the grid covers an artificial low-density atmosphere extending beyond
the stellar surface, whose rest-mass density is $ 10^{-17} $ of the
initial central rest-mass density of the star. The spectral grid of
the metric solver is split into 3 radial domains with 33 radial and 17
angular collocation points each. The innermost radial domain (or
nucleus) stretches from the coordinate origin to half the stellar
equatorial radius, followed by the second radial domain which extends
out to the outer boundary of the finite difference grid. The third
domain uses a compactified radial coordinate and reaches out to
spatial infinity. The metric equations~(\ref{eq:cfc_metric_equations})
can therefore be numerically integrated out to spacelike infinity, and
all noncompact support source terms in these equations can be
consistently handled in a non-approximative way. For further details
about the grid setup and particularly the relevance of a compactified
radial spectral grid for an accurate evolution of rotating equilibrium
models, we refer to \citet{dimmelmeier_05_a}.

Even when using spectral methods the calculation of the spacetime
metric is computationally expensive. Hence, in our simulations the
metric is updated only once every 50 hydrodynamic time steps during
evolution (which corresponds to a time interval of
$ 10^{-3} \mathrm{\ ms} $) and extrapolated in between. The
suitability of this procedure is tested and discussed in detail in
\citet{dimmelmeier_02_a}. We also note that convergence tests with
different grid resolution have been performed to ascertain that the
regular grid resolution specified above is appropriate for our
simulations.


\subsection{Fluid perturbations for exciting the pulsations}
\label{subsection:perturbations}

To excite specific eigenmodes of oscillation in the stellar models we
perturb selected equilibrium variables before starting the
evolution. In the absence of the true eigenfunction of a given mode,
each perturbation is selected so as to mimic the angular dependence of
the eigenfunction of the corresponding mode of a slowly rotating
Newtonian star. Usually, this ensures that the chosen mode will
dominate the time evolution at least for the slower rotating
models. However, since the perturbation is not exact, additional
pulsation modes will be excited, especially for rapidly rotating
models (see Section~\ref{section:recycling}).

For the $ l = 0 $ modes we use as trial eigenfunction a perturbation
of the radial component of the covariant three-velocity in the form
\begin{equation}
  v_r = a \sin \left( \pi \frac{r}{r_\mathrm{s}(\theta)} \right),
  \label{eq:perturbation_l=0}
\end{equation}
where $ r_\mathrm{s} (\theta) $ is the coordinate radius of the
surface of the star. The constant $ a $ is the amplitude of the
perturbation, for which we choose the value $ -0.005 $ (in units of
$ c $).

The $ l = 2 $ modes are excited by perturbing the $ \theta $-component
of the covariant three-velocity as follows:
\begin{equation}
  v_\theta = a \sin \left( \pi \frac{r}{r_\mathrm{s}(\theta)} \right)
  \sin \theta \, \cos \theta,
  \label{eq:perturbation_l=2}
\end{equation}
where we set $ a $ to 0.01 \citep[see][for more details about
perturbations of this form]{font_01_a}.

Correspondingly, the $ l = 4 $ perturbations are excited using the
following perturbation:
\begin{equation}
  v_\theta = a \sin \left( \pi \frac{r}{r_\mathrm{s}(\theta)} \right)
  \sin \theta \, \cos \theta \left( 3 - 7 \cos^2 \theta \right).
  \label{eq:perturbation_l=4}
\end{equation}

The time series of the evolved perturbations are Fourier analysed
after an evolution time of $ 20 \mathrm{\ ms} $, which corresponds to
a nominal frequency resolution of $ 0.05 \mathrm{\ kHz}$. Due to
the finite evolution time and the numerical viscosity which is
damping the oscillations, the frequency peaks are not
$ \delta $-functions, but correspond to a nearly symmetric bell shape,
spreading over several frequency bins. As in SAF, we use a
second-order accurate numerical derivative formula in order to find
the location of the maximum of a frequency peak, when the peaks are
well separated. In practice, this method gives agreement with expected
frequencies which is significantly better than the nominal frequency
resolution due to the finite evolution time.

The peaks in the Fourier spectra are identified with specific
pulsation modes, starting from the nonrotating member of the sequence,
where the pulsation frequencies are known from perturbation theory. As
the rotation rate increases, it becomes gradually more difficult to
identify specific modes in the Fourier spectrum. For this reason, we
also extract the two-dimensional eigenfunction for each peak in the
Fourier spectrum and use it as an additional criterion to identify
specific modes (see Section~\ref{section:recycling}).

We also emphasize that in contrast to previous work \citep[as, e.g.,
in][and SAF]{font_01_a}, which assumed the Cowling approximation of a
fixed spacetime metric, we do not use a perturbation of the rest-mass
density for the $ l = 0 $ mode,
\begin{equation}
  \delta \rho = a \rho_\mathrm{c}
  \sin \left( \pi \frac{ r}{r_\mathrm{s}(\theta)} \right),
  \label{eq:perturbation_rho}
\end{equation}
where $ \rho_\mathrm{c} $ is the central rest-mass density of the
star. In the Cowling approximation, the oscillation amplitude
$ a_\mathrm{evol} $ of the density at a particular location in the
star during the evolution not only scales linearly with the initial
perturbation amplitude $ a $ (as expected for small $ a $), but its
value is also typically close to $ a $. In contrast to this, if the
spacetime is allowed to evolve, i.e.\ it is coupled to the evolution
of the fluid, we find that the oscillation amplitude
$ a_\mathrm{evol} $ during the evolution is significantly larger than
$ a $, usually by a factor $ \sim 5 \mbox{--} 8 $. Thus even a small
initial perturbation with an amplitude of a few per cent can excite an
oscillation with a large amplitude $ a_\mathrm{evol} $, which can
possibly violate the assumption of linearity.

We attribute this effect to the fact that the perturbation enters the
hydrodynamic evolution
equations~(\ref{eq:hydro_conservation_equation}) not only through the
state vector $ \mb{U} $, but also via the metric in the form
$ \sqrt{\gamma} $, which is proportional to $ \phi^6 $. If 
$ \mb{U} $ and (via the metric
equations~\ref{eq:cfc_metric_equations}) also $ \phi^6 $ are
affected by the initial perturbation with amplitude $ a $, the
combination $ \sqrt{\gamma} \, \mb{U} $ exhibits a much larger
oscillation amplitude than $ a $, namely
$ a_\mathrm{evol} \sim (1 + a)^6 \cdot (1 + a) - 1 = (1 + a)^7 - 1 $,
which is roughly $ 7 a $ for $ a \ll 1 $. This is also reflected by an
increase of the total mass of the star by $ (1 + a)^7 $, when applying
this artificial perturbation by adding rest-mass density to the
equilibrium model, as the rest mass integral also contains a term
$ \sqrt{\gamma} \, \rho $. As the spacetime metric adapts to a new
quasi-equilibrium state with this higher rest mass, the oscillation
frequencies are systematically shifted towards higher values than
those obtained if the approximately `mass neutral'
perturbation~(\ref{eq:perturbation_l=0}) is applied. Note that this
effect of frequency shift is negligible in the Cowling approximation
when $ \delta \rho $ is small, as there $ \sqrt{\gamma} $ is constant
in time, and thus the total rest mass of the star is affected by the
density perturbation only through $ \rho $, i.e.\ like $ (1 + a) $.


\section{Equilibrium models}
\label{section:equilibrium_models}

\begin{table*}
  \begin{minipage}{140mm}
    \centering
    \caption{Properties of the four sequences of equilibrium models. A
      is a sequence of fixed rest mass $ M_0 = 1.506 \, M_\odot $ with
      $ \hat{A} = 1 $, AU is the corresponding sequence of uniformly
      rotating models, B is a sequence of fixed central rest-mass
      density $ \rho_\mathrm{c} = 1.28 \times 10^{-3} $ with
      $ \hat{A} = 1 $, and BU is the corresponding sequence of
      uniformly rotating models. All models are relativistic
      polytropes with $ N = 1 $ and $ K = 100 $.
      $ \varepsilon_\mathrm{c} $ is the central energy density with 
      $ \varepsilon = \rho (1 + \epsilon) $, $ M $ is the
      gravitational mass, $ R $ is the circumferential stellar radius,
      and $ \Omega_\mathrm{c/e} $ are the central/equatorial angular
      velocity. All other quantities are defined in the main text.}
    \label{table:equilibrium_models}
    \begin{tabular}{@{}lccrrcccc@{}}
      \hline
      Model & $ \varepsilon_\mathrm{c} $ & $ M $ &
      \multicolumn{1}{c}{$ R $} & \multicolumn{1}{c}{$ r_\mathrm{e} $} &
      $ r_\mathrm{p} / r_\mathrm{e} $ & $ \Omega_\mathrm{c} $ &
      $ \Omega_\mathrm{e} $ & $ T / |W| $ \\
      & $ (\times 10^{-3}) $ & & & & & $ (\times 10^{-2}) $ &
      $ (\times 10^{-2}) $ & \\
      \hline
      A0  & 1.444 & 1.400 &  9.59 &  8.13 & 1.000 & 0.000 & 0.000 & 0.000 \\ [0.5ex]
      A1  & 1.300 & 1.405 & 10.01 &  8.54 & 0.930 & 2.019 & 0.759 & 0.018 \\ [0.5ex]
      A2  & 1.187 & 1.408 & 10.40 &  8.92 & 0.875 & 2.580 & 0.977 & 0.033 \\ [0.5ex]
      A3  & 1.074 & 1.410 & 10.84 &  9.35 & 0.820 & 2.944 & 1.125 & 0.049 \\ [0.5ex]
      A4  & 0.961 & 1.413 & 11.37 &  9.87 & 0.762 & 3.192 & 1.232 & 0.066 \\ [0.5ex]
      A5  & 0.848 & 1.418 & 12.01 & 10.49 & 0.703 & 3.340 & 1.303 & 0.086 \\ [0.5ex]
      A6  & 0.735 & 1.422 & 12.78 & 11.25 & 0.643 & 3.383 & 1.336 & 0.107 \\ [0.5ex]
      A7  & 0.622 & 1.427 & 13.75 & 12.21 & 0.579 & 3.339 & 1.337 & 0.131 \\ [0.5ex]
      A8  & 0.509 & 1.433 & 15.01 & 13.45 & 0.513 & 3.197 & 1.300 & 0.158 \\ [0.5ex]
      A9  & 0.396 & 1.439 & 16.70 & 15.13 & 0.444 & 2.953 & 1.223 & 0.189 \\ [0.5ex]
      A10 & 0.283 & 1.447 & 19.03 & 17.44 & 0.370 & 2.604 & 1.101 & 0.223 \\
      \hline
      AU0 & 1.444 & 1.400 &  9.59 &  8.13 & 1.000 & 0.000 & 0.000 & 0.000 \\ [0.5ex]
      AU1 & 1.300 & 1.404 & 10.19 &  8.71 & 0.919 & 1.293 & 1.293 & 0.020 \\ [0.5ex]
      AU2 & 1.187 & 1.407 & 10.79 &  9.30 & 0.852 & 1.656 & 1.656 & 0.037 \\ [0.5ex]
      AU3 & 1.074 & 1.411 & 11.56 & 10.06 & 0.780 & 1.888 & 1.888 & 0.055 \\ [0.5ex]
      AU4 & 0.961 & 1.415 & 12.65 & 11.14 & 0.698 & 2.029 & 2.029 & 0.076 \\ [0.5ex]
      AU5 & 0.863 & 1.420 & 14.94 & 13.43 & 0.575 & 2.084 & 2.084 & 0.095 \\
      \hline
      B0  & 1.444 & 1.400 &  9.59 &  8.13 & 1.000 & 0.000 & 0.000 & 0.000 \\ [0.5ex]
      B1  & 1.444 & 1.437 &  9.75 &  8.24 & 0.950 & 1.801 & 0.666 & 0.013 \\ [0.5ex]
      B2  & 1.444 & 1.478 &  9.92 &  8.36 & 0.900 & 2.574 & 0.944 & 0.026 \\ [0.5ex]
      B3  & 1.444 & 1.525 & 10.11 &  8.49 & 0.850 & 3.189 & 1.160 & 0.040 \\ [0.5ex]
      B4  & 1.444 & 1.578 & 10.31 &  8.63 & 0.800 & 3.728 & 1.342 & 0.055 \\ [0.5ex]
      B5  & 1.444 & 1.640 & 10.53 &  8.77 & 0.750 & 4.227 & 1.504 & 0.071 \\ [0.5ex]
      B6  & 1.444 & 1.713 & 10.76 &  8.91 & 0.700 & 4.707 & 1.651 & 0.087 \\ [0.5ex]
      B7  & 1.444 & 1.798 & 11.01 &  9.05 & 0.650 & 5.185 & 1.789 & 0.105 \\ [0.5ex]
      B8  & 1.444 & 1.899 & 11.26 &  9.17 & 0.600 & 5.683 & 1.921 & 0.124 \\ [0.5ex]
      B9  & 1.444 & 2.020 & 11.50 &  9.26 & 0.550 & 6.232 & 2.052 & 0.144 \\
      \hline
      BU0 & 1.444 & 1.400 &  9.59 &  8.13 & 1.000 & 0.000 & 0.000 & 0.000 \\ [0.5ex]
      BU1 & 1.444 & 1.432 &  9.83 &  8.33 & 0.950 & 1.075 & 1.075 & 0.012 \\ [0.5ex]
      BU2 & 1.444 & 1.466 & 10.11 &  8.58 & 0.900 & 1.509 & 1.509 & 0.024 \\ [0.5ex]
      BU3 & 1.444 & 1.503 & 10.42 &  8.82 & 0.850 & 1.829 & 1.829 & 0.037 \\ [0.5ex]
      BU4 & 1.444 & 1.543 & 10.78 &  9.13 & 0.800 & 2.084 & 2.084 & 0.050 \\ [0.5ex]
      BU5 & 1.444 & 1.585 & 11.20 &  9.50 & 0.750 & 2.290 & 2.290 & 0.062 \\ [0.5ex]
      BU6 & 1.444 & 1.627 & 11.69 &  9.95 & 0.700 & 2.452 & 2.452 & 0.074 \\ [0.5ex]
      BU7 & 1.444 & 1.666 & 12.30 & 10.51 & 0.650 & 2.569 & 2.569 & 0.084 \\ [0.5ex]
      BU8 & 1.444 & 1.692 & 13.07 & 11.26 & 0.600 & 2.633 & 2.633 & 0.091 \\ [0.5ex]
      BU9 & 1.444 & 1.695 & 13.44 & 11.63 & 0.580 & 2.642 & 2.642 & 0.092 \\
      \hline
    \end{tabular}
  \end{minipage}
\end{table*}

Since our focus is on the effects of rotation on pulsation modes, we
do not survey here a broad range of high-density EOSs, but choose
instead a single polytropic EOS~(\ref{eq:polytropic_eos}) with
$ N = 1 $ and $ K = 100 $. For the sake of comparison we use the same
equilibrium model sequences as SAF to which we refer the reader for
more details. Here we only give a brief overview of the basic
properties of the various sequences.

We restrict our attention to two different sequences of differentially
rotating models (sequences~A and~B) and their uniformly rotating
counterparts (sequences~AU and~BU). The equilibrium properties of all
models are summarised in Table~\ref{table:equilibrium_models}. In the
nonrotating limit all sequences end in the same nonrotating model
(thus, models~A0, AU0, B0, and~BU0 all coincide). Notice that in
Table~\ref{table:equilibrium_models} the numerical values of different
equilibrium properties are displayed as computed by SAF, using the
numerical code \texttt{rns} \citep{stergioulas_95_a}. In contrast, the
equilibrium initial data used in the actual time evolutions presented
here are obtained by a similar, but different, initial data solver,
based on the original KEH method \citep{komatsu_89_a}. An extensive
comparison between the two initial data solvers shows very good
agreement with differences being much less than the 1 per cent level
in all computed equilibrium quantities (the differences arise solely
due to the truncated computational domain of the original KEH
method). We also note that for a few of the most rapidly
differentially rotating equilibrium models used by SAF, the elliptic
solver for the spacetime evolution used here did not converge. These
models are therefore omitted in Table~\ref{table:equilibrium_models}.

The differentially rotating sequence~A and its corresponding uniformly
rotating sequence~AU are characterised by a fixed rest mass
$ M_0 = 1.506 \, M_\odot $. Along sequence~A, the degree of
differential rotation is held fixed at $ \hat{A} = 1 $, where
$ \hat{A} = A / r_\mathrm{e} $ with $ r_\mathrm{e} $ being the
equatorial coordinate radius of the star and $ A $ being the rotation
parameter as defined in~\citet{komatsu_89_a}. The values of $ M_0 $
and $ \hat{A} $ are chosen in order to represent a newly-born,
differentially rotating neutron star. The angular velocity at the
equator is roughly 1/3 to 1/2 of the central angular velocity, which
is similar to the degree of differential rotation obtained in typical
core collapse simulations \citep[see, e.g.,][]{villain_04_a}. The
fastest rotating model in sequence~A has a ratio of polar to
equatorial coordinate axis of only
$ r_\mathrm{p} / r_\mathrm{e} = 0.370 $, which translates into a
rotation rate $ T / |W| = 0.223 $. (Here $ T $ is the rotational
kinetic energy and $ |W| $ is the gravitational potential energy.) The
central rest-mass density is nearly an order of magnitude smaller than
the one of the corresponding nonrotating model, while the
circumferential radius is nearly twice as large. The uniformly
rotating sequence~AU only reaches an axis ratio of 0.575 at a rotation
rate $ T / |W| = 0.095 $, half the central rest-mass density, and a
50 per cent larger radius than the corresponding nonrotating
model. Model~AU5 is at the mass-shedding limit.

On the other hand, the differentially rotating sequence~B and its
corresponding uniformly rotating sequence~BU are characterised by a
fixed central rest-mass density
$ \rho_\mathrm{c} = 1.28 \times 10^{-3} $. In sequence~B the
differential rotation parameter is also set to $ \hat{A} = 1 $. Its
fastest rotating member has a gravitational mass
$ M = 2.02 \, M_\odot $ and an axis ratio of 0.55. Since all models in
the sequence are compact, the radius $ R $ is much smaller than along
sequence~A. The corresponding uniformly rotating sequence~BU only
reaches an axis ratio of 0.58 at the mass-shedding limit (model~BU9)
with an increase in radius $ R $ by 40 per cent. Thus, we see that
when considering a sequence of fixed central rest-mass density, the
uniformly rotating models attain a larger equatorial radius than
differentially rotating models, with the latter expanding out of the
equatorial plane, becoming torus-like.


\section{Linear pulsation modes}
\label{section:linear_pulsations}

When the pulsation amplitude is small (so that, e.g., density
variations are at the level of $ \delta \rho / \rho \sim 10^{-2} $),
the dynamics is close to linear and one can identify the
eigenfrequencies and eigenfunctions of linear quasi-normal modes from
a time evolution. In order to compute the real part of the
eigenfrequency of a pulsation mode, we Fourier-transform the time
series of the evolution of a suitable physical variable (the density
for the $ l = 0 $ modes and $ v_{\theta} $ for the $ l = 2 $
modes). Instead of examining the Fourier spectra at a few specific
points inside the star, we integrate the amplitude of the Fourier
transform along a coordinate line. For instance, in the case of
$ l = 0 $ modes we examine the integrated Fourier amplitude along
$ \theta = \pi / 2 $ (equatorial plane), while for the $ l = 2 $ modes
the integrated Fourier amplitude along a line of $ \theta = \pi / 4 $
is used. We also verify that each identified discrete mode has the
same frequency at any point inside the star in the coordinate frame.

As discussed in SAF, the trial eigenfunction used for exciting the
pulsations does not correspond exactly to a particular mode.
Therefore, additional modes apart from the main mode one wishes to
study are excited, particularly for rapidly rotating models where
rotational coupling effects are significant and higher-order coupling
terms in the mode-eigenfunctions become comparable to the dominant
term. We begin the identification of specific modes along a sequence
of equilibrium models using the known pulsation frequencies of the
nonrotating star in the sequence and by comparing the Fourier
transforms of evolved variables between subsequent models. Erroneous
mode identification could happen close to the mass-shedding limit due
to the appearance of avoided crossings between different modes (see
Section~\ref{subsection:avoided_crossings}). To avoid this we do not
rely on Fourier transforms at only a few points inside the star, but
in many cases reconstruct the whole two-dimensional eigenfunction of
each mode, using Fourier transforms at every point inside the star. At
the eigenfrequency of a specific mode, the amplitude of the Fourier
transform correlates with its eigenfunction. A change in sign in the
eigenfunction corresponds to both the real and imaginary part of the
Fourier transform going through zero. Comparing the eigenfunctions
corresponding to different peaks in the Fourier transforms allows for
an unambiguous identification of specific mode sequences.

Tables~\ref{table:frequencies_a} to~\ref{table:frequencies_a_l=4}
summarise our main results, showing the frequencies of the two
lowest-order $ l = 0 $ modes ($ F $ and $ H_1 $) and $ l = 2 $ modes 
($ {}^{2\!}f $ and $ {}^{2}p_1 $) for all four sequences as well as the
two lowest order $ l = 4 $ modes ($ {}^{4}f $ and $ {}^{4}p_1 $) for
sequence~A. In addition, frequencies of several axisymmetric inertial
modes are displayed. All of the above frequency data (except for the
$ {}^{4}f $-mode, which is omitted for reasons of clarity) are shown
as a function of $ T / |W| $ in Figs.~\ref{fig:frequencies_a_au}
and~\ref{fig:frequencies_b_bu}. Next, we first discuss some general
trends due to rotation and then present our results for each mode
sequence in more detail.

\begin{table}
  \centering
  \caption{Frequencies of the fundamental quasi-radial ($ l = 0 $)
    mode, $ F $, its first overtone, $ H_1 $, the fundamental
    quadrupole ($ l = 2 $) mode, $ {}^{2\!}f $, its first overtone,
    $ {}^{2}p_1 $, and three inertial modes, $ i_{-2} $, $ i_1 $,
    and $ i_2 $ for the sequence of differentially rotating models
    A. All frequencies are given in kHz.}
  \label{table:frequencies_a}
  \begin{tabular}{@{}l@{~~}ccccccc@{}}
    \hline
    Model & $ F $ & $ H_1 $ & $ {}^{2\!}f $ & $ {}^{2}p_1 $ &
    $ i_{-2} $ & $ i_1 $ & $ i_2 $ \\
    \hline
    A0  & 1.458 & 3.971 & 1.586 & 3.726 & 0.000 & 0.000 & 0.000 \\
    A1  & 1.400 & 3.816 & 1.577 & 3.580 & 0.302 & 0.460 & 0.596 \\
    A2  & 1.358 & 3.733 & 1.567 & 3.424 & 0.399 & 0.603 & 0.779 \\
    A3  & 1.307 & 3.664 & 1.550 & 3.237 & 0.477 & 0.711 & 0.917 \\
    A4  & 1.248 & 3.583 & 1.535 & 3.013 & 0.543 & 0.794 & 1.022 \\
    A5  & 1.184 & 3.494 & 1.513 & 2.780 & 0.603 & 0.863 & 1.108 \\
    A6  & 1.105 & 3.352 & 1.482 & 2.557 & 0.646 & 0.914 & 1.163 \\
    A7  & 1.018 & 3.360 & 1.432 & 2.315 & 0.683 & 0.946 & 1.198 \\
    A8  & 0.915 & 3.114 & 1.360 & 2.068 & 0.711 & 0.965 & 1.204 \\
    A9  & 0.809 & 2.985 & 1.264 & 1.827 & 0.723 & 0.958 & 1.180 \\
    A10 & 0.685 & 2.830 & 1.098 & 1.610 & 0.712 & 0.911 & 1.104 \\
    \hline
  \end{tabular}
\end{table}

\begin{table}
  \centering
  \caption{Same as Table~\ref{table:frequencies_a}, but for the
    sequence of uniformly rotating models~AU.}
  \label{table:frequencies_au}
  \begin{tabular}{@{}l@{~~}ccccccc@{}}
    \hline
    Model & $ F $ & $ H_1 $ & $ {}^{2\!}f $ & $ {}^{2}p_1 $ &
    $ i_{-2} $ & $ i_1 $ & $ i_2 $ \\
    \hline
    AU0 & 1.458 & 3.971 & 1.586 & 3.726 & 0.000 & 0.000 & 0.000 \\
    AU1 & 1.398 & 3.785 & 1.562 & 3.455 & 0.280 & 0.354 & 0.468 \\
    AU2 & 1.345 & 3.716 & 1.554 & 3.192 & 0.384 & 0.478 & 0.611 \\
    AU3 & 1.283 & 3.635 & 1.537 & 2.885 & 0.468 & 0.575 & 0.736 \\
    AU4 & 1.196 & 3.552 & 1.516 & 2.520 & 0.545 & 0.660 & 0.810 \\
    AU5 & 1.107 & 3.457 & 1.459 & 2.090 & 0.639 & 0.747 & 0.858 \\
    \hline
  \end{tabular}
\end{table}

\begin{table}
  \centering
  \caption{Same as Table~\ref{table:frequencies_a}, but for the
    sequence of differentially rotating models~B.}
  \label{table:frequencies_b}
  \begin{tabular}{@{}l@{~~}ccccccc@{}}
    \hline
    Model & $ F $ & $ H_1 $ & $ {}^{2\!}f $ & $ {}^{2}p_1 $ &
    $ i_{-2} $ & $ i_1 $ & $ i_2 $ \\
    \hline
    B0 & 1.458 & 3.971 & 1.586 & 3.726 & 0.000 & 0.000 & 0.000 \\
    B1 & 1.407 & 3.927 & 1.628 & 3.713 & 0.258 & 0.399 & 0.518 \\
    B2 & 1.373 & 3.927 & 1.670 & 3.666 & 0.370 & 0.567 & 0.739 \\
    B3 & 1.332 & 3.964 & 1.709 & 3.584 & 0.463 & 0.702 & 0.912 \\
    B4 & 1.287 & 4.014 & 1.747 & 3.490 & 0.544 & 0.819 & 1.061 \\
    B5 & 1.237 & 4.072 & 1.789 & 3.390 & 0.622 & 0.921 & 1.208 \\
    B6 & 1.178 & 4.118 & 1.819 & 3.280 & 0.681 & 1.015 & 1.318 \\
    B7 & 1.077 & 4.179 & 1.854 & 3.187 & 0.750 & 1.090 & 1.439 \\
    B8 & 1.080 & 4.212 & 1.900 & 3.103 & 0.827 & 1.213 & 1.562 \\
    B9 & 0.945 & 4.469 & 1.917 & 3.028 & 0.890 & 1.294 & 1.670 \\
    \hline
  \end{tabular}
\end{table}

\begin{table}
  \centering
  \caption{Same as Table~\ref{table:frequencies_a}, but for the
    sequence of uniformly rotating models~BU.}
  \label{table:frequencies_bu}
  \begin{tabular}{@{}l@{~~}ccccccc@{}}
    \hline
    Model & $ F $ & $ H_1 $ & $ {}^{2\!}f $ & $ {}^{2}p_1 $ &
    $ i_{-2} $ & $ i_1 $ & $ i_2 $ \\
    \hline
    BU0 & 1.458 & 3.971 & 1.586 & 3.726 & 0.000 & 0.000 & 0.000 \\
    BU1 & 1.413 & 3.915 & 1.611 & 3.634 & 0.224 & 0.288 & 0.385 \\
    BU2 & 1.380 & 3.907 & 1.635 & 3.516 & 0.326 & 0.413 & 0.541 \\
    BU3 & 1.343 & 3.921 & 1.669 & 3.345 & 0.408 & 0.511 & 0.658 \\
    BU4 & 1.304 & 3.950 & 1.698 & 3.200 & 0.486 & 0.601 & 0.762 \\
    BU5 & 1.281 & 3.964 & 1.714 & 3.018 & 0.552 & 0.675 & 0.847 \\
    BU6 & 1.219 & 4.010 & 1.729 & 2.859 & 0.617 & 0.743 & 0.912 \\
    BU7 & 1.207 & 4.018 & 1.720 & 2.677 & 0.680 & 0.809 & 0.969 \\
    BU8 & 1.168 & 4.030 & 1.685 & 2.512 & 0.723 & 0.868 & 1.004 \\
    BU9 & 1.169 & 4.029 & 1.679 & 2.483 & 0.737 & 0.878 & 1.004 \\
    \hline
  \end{tabular}
\end{table}

\begin{table}
  \centering
  \caption{Frequencies of the fundamental hexadecapole ($ l = 4 $) mode,
    $ {}^{4}f $, and its first overtone, $ {}^{4}p_1 $, for the sequence
    of differentially rotating models~A (left) and~B (right). All
    frequencies are given in kHz.}
  \label{table:frequencies_a_l=4}
  \begin{tabular}{@{}lccclcc@{}}
    \hline
    Model & $ {}^{4}f $ & $ {}^{4}p_1 $ & \quad \quad \quad &
    Model & $ {}^{4}f $ & $ {}^{4}p_1 $ \\ [0.4 em]
    \cline{1-3}
    \cline{5-7} \\ [-0.8 em]
    A0  & 2.440 & 4.896 & & B0 & 2.440 & 4.896 \\
    A1  & 2.370 & 4.727 & & B1 & 2.453 & 4.877 \\
    A2  & 2.300 & 4.600 & & B2 & 2.468 & 4.855 \\
    A3  & 2.223 & 4.372 & & B3 & 2.486 & 4.827 \\
    A4  & 2.130 & 4.130 & & B4 & 2.500 & 4.781 \\
    A5  & 2.028 & 3.864 & & B5 & 2.504 & 4.740 \\
    A6  & 1.910 & 3.617 & & B6 & 2.491 & 4.643 \\
    A7  & 1.780 & 3.116 & & B7 & 2.499 & 4.556 \\
    A8  & 1.630 & 2.790 & & B8 & 2.501 & 4.506 \\
    A9  & 1.480 & 2.430 & & B9 & 2.493 & 4.164 \\
    A10 & 1.330 & 2.028 \\
    \hline
  \end{tabular}
\end{table}


\subsection{General trends of rotational effects}
\label{subsection:rotation_trends}

It is well known that the frequencies of the fundamental $ l = 0 $ and
$ l = 2 $ polar modes of oscillation depend mainly on the central
density of a star, or, equivalently, on the compactness $ M / R $
\citep[see, e.g.,][]{hartle_75_a}. The sequences~A and~AU of fixed
rest mass $ M_0 = 1.506 \, M_\odot $ start with a nonrotating model
with compactness $ M / R = 0.15 $ and terminate at models with much
smaller compactness ($ M / R = 0.076 $ for sequence~A and
$ M / R = 0.095 $ for sequence~AU). Based on this significant decrease
of the compactness along the fixed-rest-mass sequences, one expects a
corresponding decrease in the frequencies of the fundamental modes
(and a similar tendency for the first overtones). In contrast, along
sequences~B and~BU, where the central density is fixed, the
compactness varies much less than for sequences~A and~AU. In fact, for
sequence~B, the compactness even somewhat \emph{increases}. One
therefore expects a weaker dependence of the pulsation frequencies on
rotation for the sequences of fixed central density. The above
expectations have already been verified qualitatively in the Cowling
approximation by SAF.

\begin{figure}
  \includegraphics[width=84mm]{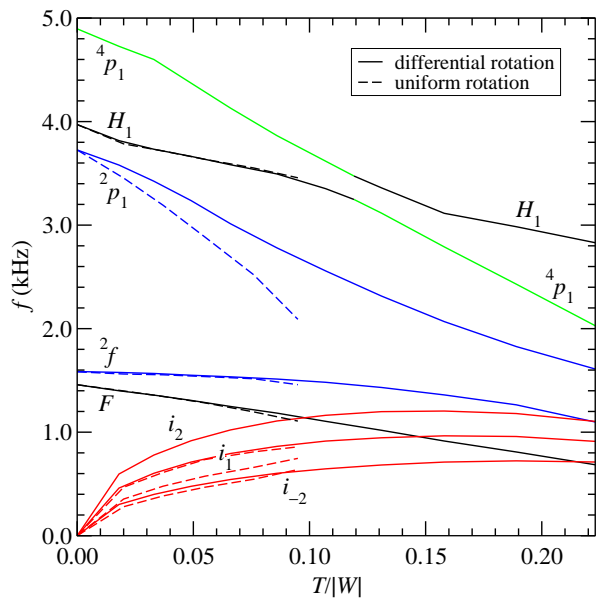}
  \caption{Frequencies of various modes for sequences~A (solid
    lines) and~AU (dashed lines). Note the avoided crossing between
    the $ H_1 $ and the $ {}^{4}p_1 $-mode.}
  \label{fig:frequencies_a_au}
\end{figure}

In slowly rotating stars, the frequencies of all inertial modes
increase linearly with increasing $ T / |W| $
\citep[see][]{friedman_01_a, lockitch_03_a}. At higher rotation
rates, higher-order rotational terms can modify this behavior. For
uniformly rotating stars, the expectation is that the inertial mode
frequency still increases up to the mass-shedding limit. As clearly
visible for the rapidly rotating models of sequence~A in
Fig.~\ref{fig:frequencies_a_au}, this general expectation is no longer
valid for differentially rotating stars (for a more detailed
discussion, see Section~\ref{subsection:inertial_modes}).

\begin{figure}
  \includegraphics[width=84mm]{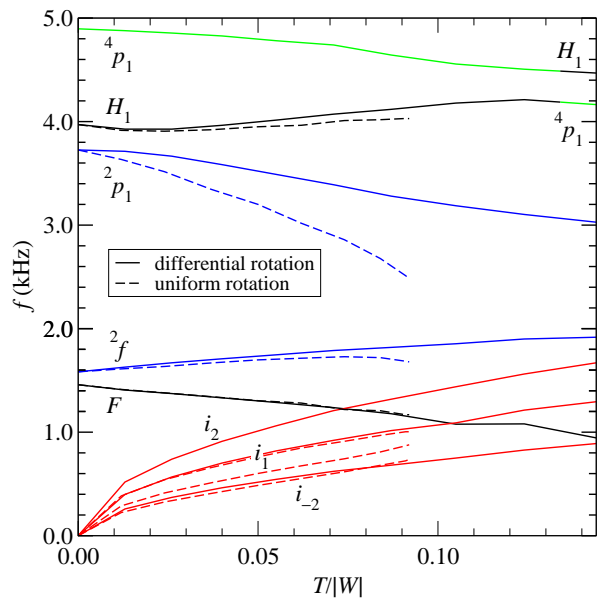}
  \caption{Same as Fig.~\ref{fig:frequencies_a_au}, but for the sequences~B and
    BU.}
  \label{fig:frequencies_b_bu}
\end{figure}

Due to differential rotation, the outer layers of the star rotate
slower and the equatorial radius is smaller compared to a uniformly
rotating model of same $ T / |W| $. This leads to a smaller
sound-crossing time and correspondingly higher fundamental mode
frequencies for the differentially rotating models. This explains why
the curves for the fundamental mode frequencies of the $ F $ and
$ {}^{2\!}f $-mode of sequence~A in Fig.~\ref{fig:frequencies_a_au} have
smaller slopes than those corresponding to sequence~AU. This behavior
was already found for the same model sequence in the Cowling
approximation in the work of SAF.

In general, higher order or large $ l $ modes are affected more strongly
by rotation than lower order or smaller $ l $ modes. At large rotation
rates this can lead to \emph{avoided crossings} between mode
sequences, where modes can exchange the character of their
eigenfunctions. These avoided crossings are already known to
exist from perturbative studies of axisymmetric modes in rotating
Newtonian stars~\citep{clement_86_a} and for relativistic quasi-radial
modes~\citep{yoshida_01_a}. It is not trivial to decide how to 
label the mode sequences after an avoided crossing. The decisive
criterion for labeling a pulsation mode is \emph{not} the continuity
of the eigenfrequency along a mode sequence. More important is the
character of the oscillation, i.e.\ the eigenfunction. One must
therefore examine the eigenfunctions of two pulsations before and
after an avoided crossing. Then the character of the modes after the
crossing can be determined according to which modes (of those before
the crossing) they resemble. Thus, continuity of eigenfunctions is
preferred over continuity of eigenfrequencies in labeling modes.


\subsection{Quasi-radial (\boldmath $ l = 0 $) modes and avoided
  crossings}
\label{subsection:avoided_crossings}

The computed frequencies for the fundamental quasi-radial $ l = 0 $
mode $ F $ and its first overtone $ H_1 $ for the
fixed rest mass sequences~A and~AU are displayed in
Tables~\ref{table:frequencies_a} and~\ref{table:frequencies_au} and
plotted in Fig.~\ref{fig:frequencies_a_au}. Along the uniformly
rotating sequence~AU, there is a decrease in the frequency of the
fundamental quasi-radial mode, since the central density of the star
decreases with increasing rotation rate. For Newtonian nonrotating
polytropic models the frequency of the fundamental quasi-radial mode
is proportional to the square root of the average density. Even though
we do not compute an average density for the rapidly rotating models
of sequences~A and~AU, we notice that the decrease in the frequency of
the $ F $-mode from $ 1.458 \mathrm{\ kHz} $ to $ 1.107 \mathrm{\ kHz} $
along sequence~AU follows closely the decrease in central energy
density $ \varepsilon_\mathrm{c} $. Along the differentially rotating
sequence A, the frequency of the $ F $-mode is further decreasing,
reaching a very low value of $ 685 \mathrm{\ Hz} $ for the most
rapidly rotating model. Remarkably, when one compares models of same
$ T / |W| $ along the two sequences~A and~AU, the frequency of the
$ F $-mode is insensitive to the degree of differential rotation.

The first overtone $ H_1 $ also decreases along the uniformly rotating
sequence~AU, but by less than the near
$ \sqrt{\varepsilon_\mathrm{c}} $-dependence of the $ F $-mode. Along
the differentially rotating sequence A, the $ H_1 $-mode shows a
similar insensitivity to the degree of differential rotation as the
$ F $-mode, when comparing models of same $ T / |W| $, up to rotation
rates of $ T / |W| \sim 0.1 $. For larger rotation rates, however, an
extended \emph{avoided crossing} with the $ {}^{4}p_1 $-mode takes
place. Such avoided crossings exist, because the various mode
sequences are affected to a different degree by rotation. In
particular, the frequencies of higher $ l $ modes tend to decrease
faster with increasing $ T / |W| $ than the frequencies of lower $ l $
modes, leading to approaching mode-sequences. However, mode-sequences
that contain similar terms in their eigenfunction expansions are not
allowed to cross, even in the linear
approximation~\citep{clement_86_a, yoshida_01_a}. Instead, two
continuous sequences of pulsation frequencies avoid to cross, as shown
in Fig.~\ref{fig:frequencies_a_au}.

The character of the eigenfunctions along these continuous sequences
is \emph{exchanged} at an avoided crossing. As a result, the
lower-frequency part of the continuous sequence starting as the
$ H_1 $-mode in the nonrotating limit becomes the $ {}^{4}p_1 $-mode
at large $ T / |W| $. Correspondingly, the lower-frequency part of the
continuous sequence which starts as the $ {}^{4}p_1 $-mode in the
nonrotating limit, becomes the $ H_1 $-mode at large $ T / |W| $. At
the avoided crossing, the character of the eigenfunctions of both
modes is a mixture of the eigenfunctions of the two modes before the
avoided crossing. We have determined the correct labeling of the
sequences after the avoided crossing by carefully comparing their
two-dimensional eigenfunctions (see Section~\ref{section:recycling}
for a discussion of how the eigenfunctions are obtained from our
time-evolutions). The particular avoided crossing between the $ H_1 $
and $ {}^{4}p_1 $-modes was also observed along a sequence of rotating
models in the relativistic Cowling approximation \citep{yoshida_01_a}.

Along the fixed central density sequences~B and~BU (see
Tables~\ref{table:frequencies_b} and~\ref{table:frequencies_bu} and
Fig.~\ref{fig:frequencies_b_bu}), the frequency of the $ F $-mode
decreases, because, even though the central density stays fixed,
rotational effects still increase the radius of the star, so that the
sound crossing time in the equatorial region increases. This seems to
have a monotonous influence on the frequency of the $ F $-mode, which
is again insensitive to the degree of differential rotation. The
frequency of the $ H_1 $-overtone, on the other hand, first decreases
somewhat with rotation, but then increases again, eventually
surpassing its value of $ 3.971 \mathrm{\ kHz} $ in the nonrotating
limit. Since the $ H_1 $-mode has a node in its eigenfunction in the
nonrotating limit, it is more sensitive to rotational effects (in
comparison to the fundamental $ F $-mode). These rotational effects
have a different influence near the symmetry axis than near the
equator. The changing dependence of the $ H_1 $-mode frequency as the
star becomes more flattened reflects this sensitivity. In the case of
sequence~B, the avoided crossing between the $ H_1 $-mode and the
$ {}^{4}p_1 $-mode happens at the largest rotation rates along this
sequence. Consequently, the fastest rotating model~B9 is in the
avoided crossing region, where the mode eigenfunctions are mixed.


\subsection{Quadrupole (\boldmath $ l = 2 $) modes}
\label{subsection:quadrupole_modes}

The computed frequencies for the fundamental quadrupole $ l = 2 $ mode
$ {}^{2\!}f $ and its first overtone $ {}^{2}p_1 $ for the fixed rest
mass sequences~AU and~A are displayed in
Tables~\ref{table:frequencies_au} and~\ref{table:frequencies_a} and
plotted in Fig.~\ref{fig:frequencies_a_au}. Along the uniformly
rotating sequence~AU, there is only a small decrease in the frequency
of the fundamental $ {}^{2\!}f $-mode, from $1.586 \mathrm{\ kHz}$ to
$ 1.459 \mathrm{\ kHz} $ at the mass-shedding limit. However, for the
differentially rotating sequence~A, the rate of decrease gets stronger
for models with very high $ T / |W| $, and the frequency of the
fundamental $ {}^{2\!}f $-mode becomes as small as
$ 1.098 \mathrm{\ kHz} $.

The first overtone $ {}^{2}p_1 $ shows a much stronger decrease in
frequency with increasing rotation rates. Starting from
$ 3.726 \mathrm{\ kHz} $ at the nonrotating limit, along sequence~AU
its frequency becomes $ 2.090 \mathrm{\ kHz} $ at mass-shedding, while
it has a value of only $ 1.610 \mathrm{\ kHz} $ for the fastest
differentially rotating model along sequence~A. A striking difference
compared to all other modes studied here is that the frequency of the
$ {}^{2}p_1 $-mode is indeed sensitive to the degree of differential
rotation, as is evident from Figs.~\ref{fig:frequencies_a_au}
and~\ref{fig:frequencies_b_bu}.

Along the fixed central density sequences~B and~BU the
$ {}^{2\!}f $-mode shows the opposite tendency compared to its behavior
along the sequences~A and~AU, with its frequency increasing to
$ 1.679 \mathrm{\ kHz} $ and $ 1.917 \mathrm{\ kHz} $, respectively
(see Tables~\ref{table:frequencies_b} and~\ref{table:frequencies_bu}
and Fig.~\ref{fig:frequencies_b_bu}). In contrast, the frequency of
the $ {}^{2}p_1 $-mode is still decreasing (although not as
drastically as for sequences~A and~AU) with increasing rotation rates,
reaching frequencies of $ 3.028 \mathrm{\ kHz} $ at mass-shedding
along sequence~B and $ 2.483 \mathrm{\ kHz} $ along sequence~BU. The
$ {}^{2}p_1$-mode appears to be even more sensitive to the degree of
differential rotation along the fixed central density sequence~B than
along the fixed rest mass sequence~A.


\subsection{Inertial modes}
\label{subsection:inertial_modes}

In our simulations we observe a large number of inertial modes, 
which are supported by the Coriolis force and become degenerate
at zero frequency in nonrotating stars. From linear perturbation
theory we expect that there exists an infinite number of inertial
modes in a finite frequency range, which corresponds to 0 to
$ 2 \, \Omega $ for uniformly rotating Newtonian stars \citep[see,
e.g.,][]{lockitch_99_a}. In spite of this, we do not find evidence for
the excitation of an arbitrary number of inertial modes, but only a
few specific inertial modes are predominantly excited. Since we do not
make use of the eigenfunction recycling method to excite inertial
modes (see Section~\ref{section:recycling}), they are excited as
by-products of the excitation of other modes. Hence, inertial modes
can be excited either due to the non-exact nature of the initial trial
eigenfunctions used to perturb the initial model (as described in
Section~\ref{subsection:perturbations}) or due to non-linear couplings
with linear polar modes (see discussion in
Section~\ref{subsection:nonlinear_couplings}). The fact that the
amplitude of the observed inertial modes, with our choice of trial
eigenfunctions, scales non-linearly with increasing amplitude, points
to non-linear couplings as a possible origin, at least for most of the
inertial modes we observe.

Examining the eigenfunctions of the excited inertial modes we notice
that the mode with the usually highest PSD in a Fourier transform has
only one node in the eigenfunction for $ v_{\theta}$, while other
modes at larger or smaller frequency than this `fundamental'
inertial mode have a larger number of nodes (increasing as the
absolute frequency difference from the `fundamental' inertial mode
increases). Out of the many excited inertial modes, we choose to
display in Tables~\ref{table:frequencies_a}
to~\ref{table:frequencies_bu} and Figs.~\ref{fig:frequencies_a_au}
and~\ref{fig:frequencies_b_bu} the `fundamental' inertial mode with
only one node in $ v_{\theta} $, which we call $ i_1 $-mode, and the
two modes with two nodes in the eigenfunction of $ v_{\theta} $ on
both sides of the $ i_1 $-mode in the frequency domain, which we
denote the $ i_{-2} $-mode and $ i_2 $-mode, respectively.

\begin{figure}
  \includegraphics[width=84mm]{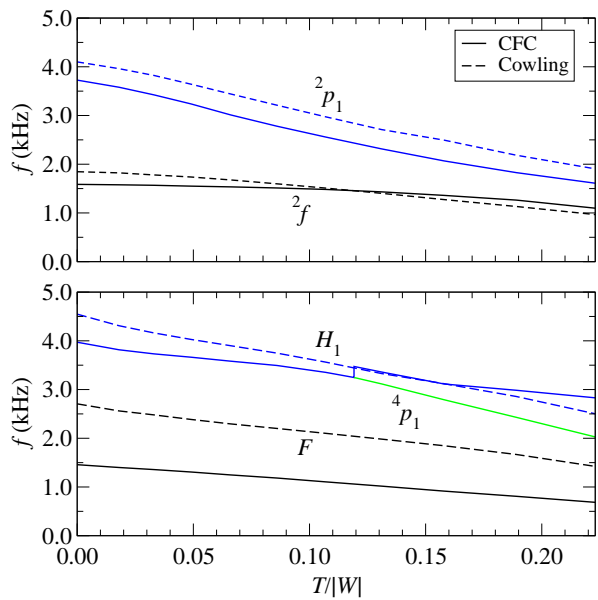}
  \caption{Comparison of the frequencies for the $ {}^{2\!}f $ and
    $ {}^{2}p_1 $-mode (upper panel) and for the $ F $ and
    $ H_1 $-mode (lower panel) in CFC (solid lines) and in the Cowling
    approximation (dashed lines) for the sequence of differentially
    rotating models~A. In the lower panel we also show the frequencies
    of the $ {}^{4}p_1 $-mode after the avoided crossing with the
    $ H_1 $-mode (green line).}
  \label{fig:cfc_comparison_a}
\end{figure}

Notice that the number of nodes is determined at slow rotation rates
and can change at very high rotation rates. Note also that inertial
modes come in two flavours (polar-led and axial-led) and are hybrid
modes in the sense that they do not reduce to either a pure polar or a
pure axial mode in the nonrotating limit
\citep[see][]{lockitch_99_a}. A proper classification scheme for
inertial modes, based on the dominant components in their
eigenfunctions, is already in use \citep{lockitch_99_a,
  friedman_01_a, lockitch_03_a}. However, comparing our numerically
calculated eigenfunctions with eigenfunctions obtained from linear
perturbation theory with more simplifying assumptions (weak gravity or
uniform density) is beyond the scope of the present paper and will be
addressed in a separate publication. Thus, we use our own naming
scheme for inertial modes simply for convenience, until proper
identification is established in the future.

\begin{figure}
  \includegraphics[width=84mm]{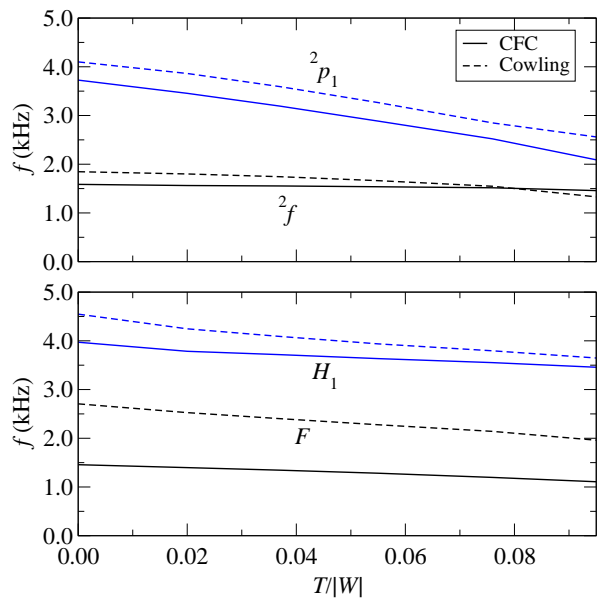}
  \caption{Same as Fig.~\ref{fig:cfc_comparison_a}, but for the
    sequence of uniformly rotating models~AU.}
  \label{fig:cfc_comparison_au}
\end{figure}

Along sequence~AU the frequencies of the three inertial modes
increases monotonically with rotation. However, along sequence~A the
frequencies of the three inertial modes reach a maximum value for the
rotation parameter $ T / |W| $ being in the range 0.15 to 0.19,
depending on the specific mode. This is due to the fact that, even
though $ T / |W| $ increases monotonically along sequence~A, both the
angular velocity at the centre and the angular velocity at the surface
reach maximum values along this sequence, as can be seen in
Table~\ref{table:equilibrium_models}. On the other hand, along
sequences~BU and~B the frequencies of the three inertial modes again
increase monotonically, as does the angular velocity at the centre and
at the surface.


\subsection{Comparison to the Cowling approximation}
\label{subsection:comparison_to_cowling}

\begin{figure}
  \includegraphics[width=84mm]{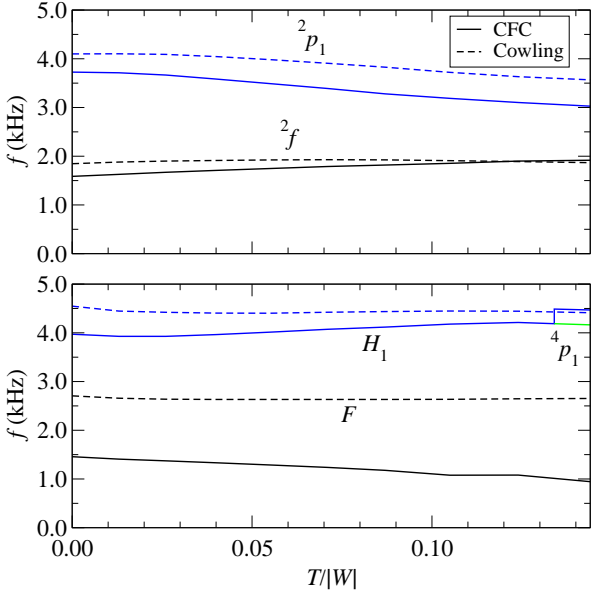}
  \caption{Same as Fig.~\ref{fig:cfc_comparison_a}, but for the
    sequence of differentially rotating models~B.}
  \label{fig:cfc_comparison_b}
\end{figure}

\begin{figure}
  \includegraphics[width=84mm]{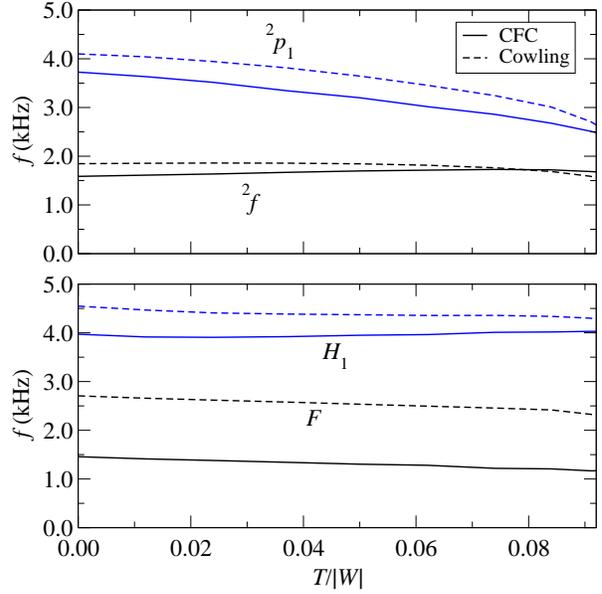}
  \caption{Same as Fig.~\ref{fig:cfc_comparison_a}, but for the
    sequence of uniformly rotating models~BU.}
  \label{fig:cfc_comparison_bu}
\end{figure}

When comparing the dependence of the mode frequency on $ T / |W| $
plotted in Figs.~\ref{fig:cfc_comparison_a}
to~\ref{fig:cfc_comparison_bu} for the $ F $, $ H $, $ {}^{2\!}f $, and
$ {}^{2}p_1 $-modes obtained with the CFC approximation for the
spacetime evolution against previous results in the Cowling
approximation by SAF, several qualitative observations can be
made. Apparently, for all sequences there is a large discrepancy in
the values of the frequencies of the fundamental quasi-radial
$ F $-mode in the CFC simulations compared to the ones in the Cowling
approximation at all rotation rates. This is also quantitatively
evident in the relative differences of frequencies between the CFC and
Cowling simulations of up to a factor $ \sim 2 $, listed for the
nonrotating model (A0/AU0/B0/BU0) and the most rapidly rotating model
of each sequence (A10, AU5, B9, BU9) in
Table~\ref{table:comparsion_cfc_cowling}. We notice that these
findings are consistent with those of \citet{yoshida_97_a} for
nonrotating stars. In terms of absolute differences, for the $ F
$-mode the Cowling approximation fails by far to predict the correct
mode frequencies, as it yields values which are typically
$ \gtrsim 1 \mathrm{\ kHz} $ too high. While in sequences~A, AU,
and~BU the frequency curve in the Cowling approximation at least
correctly captures the decline with increasing rotation rate, it
cannot reproduce the similar behavior in sequence~B.

We emphasize that neither in sequence~A nor~B we find any evidence for
a splitting of the $ F $-mode, a phenomenon that was noticed for those
sequences in the Cowling approximation by SAF. This apparently
confirms the possible explanation for the $ F $-mode splitting as an
artifact of the Cowling approximation offered in that work. In
Figs.~\ref{fig:cfc_comparison_a} to~\ref{fig:cfc_comparison_bu} and
Table~\ref{table:comparsion_cfc_cowling} we use the frequencies of the
regular $ F $-mode rather than those of the $ F_\mathrm{II} $-mode
of SAF, as the eigenfunctions of the latter mode do not possess the
typical characteristics of a genuine $ F $-mode (and, in addition, 
their amplitude of the frequency peak in FFTs is smaller than for
the $ F $-mode). 

Figs.~\ref{fig:cfc_comparison_a} to~\ref{fig:cfc_comparison_bu}
demonstrate that for the frequencies of the fundamental quadrupole
$ {}^{2\!}f $-mode and the $ H_1 $ and $ {}^{2}p_1 $-modes, the
discrepancies between the CFC simulations and the Cowling
approximation are much less severe (see also
Table~\ref{table:comparsion_cfc_cowling}). In the case of the
$ H_1 $-mode and the $ {}^{2}p_1 $-mode, the curve in the Cowling
approximation follows our CFC results reasonably close, while we
observe a crossing of the frequency curves for the $ {}^{2\!}f $-mode
beyond medium rotation rates in all sequences. Nevertheless, the
$ {}^{2\!}f $-mode frequencies still agree well quantitatively. Note
however that avoided crossings of the $ H_1 $-mode and the
$ {}^{4}p_1 $-mode (as in sequences~A and~B) complicates a
comparison. If the continuous frequency curve of the $ H_1 $-mode is
followed, the $ H_1 $-mode takes over the characteristics of the
$ {}^{4}p_1 $-mode at the avoided crossing. Thus, if \emph{continuous}
frequency curves are compared, a possible change of the mode labeling
must be taken into account. Although the simulations in the Cowling
approximation reproduce the $ H_1 $, $ {}^{2\!}f $, and
$ {}^{2}p_1 $-mode frequencies from the CFC simulations fairly close
for most sequences, which supports the idea of establishing empirical
relations for predicting the correct frequencies from models evolved
in the Cowling approximation, such relations should be used with
caution (see discussion below). 

\begin{table}
  \centering
  \caption{Relative difference in per cent in the frequencies of the
    $ F $, $ H_1 $, $ {}^{2\!}f $, and $ {}^{2}p_1 $-mode for the
    nonrotating model and the most rapidly rotating models of
    sequence~A, AU, B, and~BU in the Cowling approximation with
    respect to the CFC simulations. A $ + $~sign ($ - $~sign)
    indicates an overestimation (underestimation) of the CFC frequency
    in the Cowling approximation.}
  \label{table:comparsion_cfc_cowling}
  \begin{tabular}{@{}l@{~~}rrrr@{}}
    \hline
    Model & $ F $ & $ H_1 $ & $ {}^{2\!}f $ & $ {}^{2}p_1 $ \\
    \hline
    A0/AU0/B0/BU0  & $  +86 $ & $ +15 $ & $ +16 $ & $ +10 $ \\
    A10            & $ +108 $ & $ -12 $ & $ -12 $ & $ +18 $ \\
    AU5            & $  +77 $ & $  +5 $ & $  -9 $ & $ +22 $ \\
    B9             & $  +81 $ & $  -1 $ & $  -3 $ & $ +18 $ \\
    BU9            & $  +98 $ & $  +6 $ & $  -7 $ & $  +6 $ \\
    \hline
  \end{tabular}
\end{table}

\begin{figure}
  \includegraphics[width=84mm]{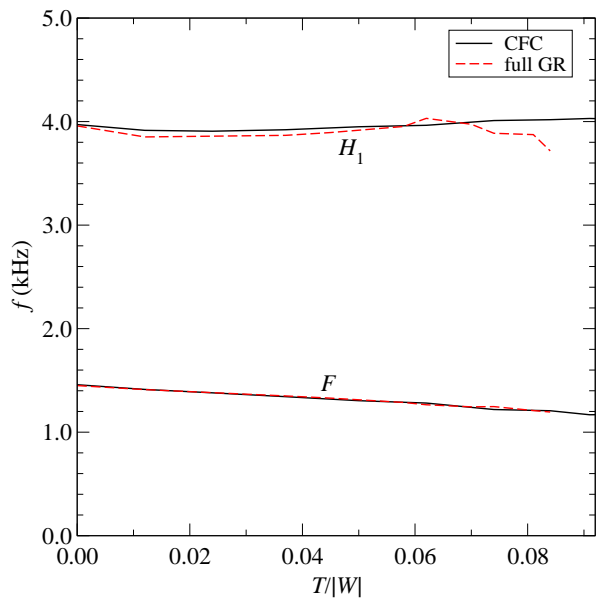}
  \caption{Comparison of the frequencies for the $ F $ and
    $ H_1 $-mode in CFC (black solid lines) and full general
    relativity (red dashed lines) for sequence~BU.}
  \label{fig:gr_comparison_bu}
\end{figure}

For sequence~BU a comparison of models evolved using a coupled
evolution of hydrodynamics and spacetime, with no approximation for
the gravitational field equations, against the Cowling approximation
was presented by \citet{font_02_a}\footnote{Note that in that work the
  sequence already ended at our model BU7 with an axis ratio
  $ r_\mathrm{p} / r_\mathrm{e} = 0.650 $, while several additional
  models with intermediate rotation rates were also evolved.}. 
Comparing these results with our new simulations using the CFC
approximation as presented in Fig.~\ref{fig:gr_comparison_bu} shows
excellent agreement for the frequencies of the $ F $-mode. This is
another important confirmation of the validity of the results obtained
with our code, considering in particular the differences in coordinate
choice, grid resolution, evolution time, and assumption of CFC
compared to the code employed in \citet{font_02_a}\footnote{A similar
  comparison for sequence~AU yields equally good results
  \citep{bernuzzi_06_a}.}.

In the case of the $ H_1 $-mode we also find rather good agreement,
with increasing mismatch as the models approach the mass-shedding
limit. The oscillations in the $ H_1 $-mode frequency at large
rotation rates observed in the fully relativistic simulations by
\citet{font_02_a} were interpreted as possible effects of avoided
crossings. In contrast to this, in our simulations we find neither
these oscillations nor signs of avoided crossing of the $ H_1 $-mode
with other modes in sequences~AU and~BU (see
Fig.~\ref{fig:gr_comparison_bu}, and also
Figs.~\ref{fig:frequencies_a_au} and~\ref{fig:frequencies_b_bu}). For
sequences~A and~B, where there is clear evidence for avoided crossing,
this happens at larger rotation rates. We thus conclude that the
oscillations visible in Fig.~\ref{fig:gr_comparison_bu} (upper red
dashed line) are an artifact which can be explained by the effectively
lower grid resolution and shorter evolution time compared to the
current simulations, as this increases the error in the numerical
extraction of the mode frequency particularly for higher order modes.

When comparing the frequencies for sequence~BU from simulations with
a dynamical spacetime to the ones in the Cowling approximation,
\citet{font_02_a} also observed that the frequencies for the
$ F $-mode (and, to a lesser degree, also the ones for the
$ H_1 $-mode) show a similar dependence on the rotation rate in both
cases. From the approximate constancy of the difference between the
$ F $-mode frequency and the corresponding result in the Cowling
approximation, they constructed an approximate empirical relation for
calculating the frequency of the $ F $-mode for arbitrary rotation
rates. This relation thus depends on the $ F $-mode frequency of the
nonrotating model~BU0 obtained in a dynamical spacetime evolution
and the variation of the $ F $-mode frequencies with increasing
rotation in the Cowling approximation. It yields the correct
frequencies with an accuracy of better than 2 per cent for the most
rapidly rotating in their model series of sequence~BU (our model~BU7).

Unfortunately our results clearly suggest that it is not possible to
straightforwardly apply this simple method to set up analogous
relations for sequences~A, AU, and~B, as shown in
Figs.~\ref{fig:cfc_comparison_a} to~\ref{fig:cfc_comparison_bu}. Only
for sequence~BU, the $ F $-mode frequency exhibits a nearly constant
difference between the CFC and Cowling simulations irrespective of the
rotation rate. Moreover, the frequencies of all other modes displayed
in these figures also do not fulfill such a relation, except maybe in
the case of the $ {}^{2}p_1 $-mode for sequences~A and~AU.

Lacking results from simulations involving evolution of the
spacetime, SAF proposed an approximate relation for the $ F $ and
$ {}^{2\!}f $-mode frequencies of sequence~A similar to the one in
\citet{font_02_a}. Based on the work by \citet{yoshida_97_a} and
\citet{yoshida_01_a}, an empirical relation was derived using
information from the compactness of a model, and assuming nearly
linear scaling of the frequencies with increasing rotation rate. When
applying this relation to our model~A10, we find a significant
discrepancy between the predicted frequencies
($ f_F = 0.882 \mathrm{\ kHz} $ and
$ f_{H_1} = 0.757 \mathrm{\ kHz} $) and the ones extracted from the
actual numerical simulations ($ f_F = 0.685 \mathrm{\ kHz} $ and
$ f_{H_1} = 1.098 \mathrm{\ kHz} $), with relative differences of 29
and 31 per cent for the $ F $ and $ {}^{2\!}f $-mode frequencies,
respectively. Although their assumption of nearly linear scaling of
$ f_F $ and $ f_{H_1} $ with the rotation rate is confirmed by our
results (see Figs.~\ref{fig:cfc_comparison_a}
to~\ref{fig:cfc_comparison_bu}), this error is much larger than the
predicted uncertainty of the relation of a few per cent.

Owing to this, we refrain from establishing similar relations for
other mode frequencies, even in cases where we also observe such
nearly linear scaling of the frequency with rotation rate in CFC
simulations and/or constant difference between the frequency of a
specific mode in CFC and Cowling. We rather suggest that except in
special cases, (the computationally expensive) simulations of rotating
stars in which the spacetime and the hydrodynamics are coupled cannot
be replaced by a combination of simulations in the Cowling
approximation and simple empirical relations.


\section{Eigenfunction recycling}
\label{section:recycling}

\begin{figure}
  \includegraphics[width=84mm]{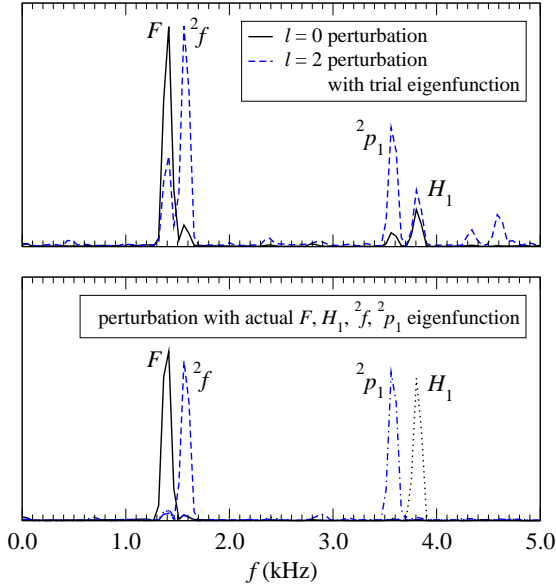}
  \caption{Fourier transform of the evolution of the radial profile of
    the rest-mass density $ \rho $ along the equatorial
    plane for model A1. For the two simulations in the upper panel an
    $ l = 0 $ and $ l = 2 $ trial eigenfunction is used as
    perturbation, respectively, while the four recycling runs in the
    lower panel are excited with recycled eigenfunctions. The scaling
    of the ordinate is linear, and the power spectra in the upper
    panel are scaled to obtain the same strength of the $ F $-mode.}
  \label{fig:recycling}
\end{figure}

As pointed out in Section~\ref{subsection:perturbations}, initial
perturbations of the form given by Eqs.~(\ref{eq:perturbation_l=0},
\ref{eq:perturbation_l=2}) excite not only one desired specific
eigenmode, but additional oscillation modes as well. This is clearly
visible in the power spectrum of model~A1 displayed in the upper panel
of Fig.~\ref{fig:recycling}. While the fundamental modes $ F $ and
$ {}^{2\!}f $ clearly dominate the spectrum, the overtones $ H_1 $ and
$ {}^{2}p_1 $ along with several other modes are also significantly
excited by a perturbation using the $ l = 0 $ and $ l = 2 $ trial
eigenfunction of the initial velocity components $ v_r $ and
$ v_{\theta} $, respectively.

This consequence of using trial eigenfunctions can be avoided by
performing an additional `recycling' run. For this the \emph{actual}
two-dimensional eigenfunctions of the $ v_r $ and $ v_{\theta} $
velocity components at a selected mode frequency are extracted from
the original simulation which was perturbed by a trial
eigenfunction. These eigenfunctions are then applied with appropriate
amplitude as initial perturbation to a second simulation of the same
stellar equilibrium model. In this case, other modes than the selected
one are strongly suppressed, as shown for model~A1 in the lower panel
of Fig.~\ref{fig:recycling}. Here eigenfunctions of the $ F $ and
$ H_1 $-mode, and the $ {}^{2\!}f $ and $ {}^{2}p_1 $-mode are extracted
from a previous simulation perturbed with $ l = 0 $ and $ l = 2 $
trial eigenfunctions, respectively. These are then used as initial
perturbations for four recycling runs with relative amplitudes with
respect to the amplitude of the $ F $ eigenfunction of 5.9, 3.8, and
7.0 for the $ H_1 $, $ {}^{2\!}f $, and $ {}^{2}p_1 $ eigenfunction,
respectively, in order to arrive at approximately equal strengths of
the dominant modes in the power spectrum.

We note that the choice of using $ v_r $ and $ v_\theta $
eigenfunctions as initial perturbations for doing the recycling
simulation is not arbitrary. Naively, one would think to use all
four evolved variables $ \rho $, $ v_r $, $ v_\theta $, and
$ v_\varphi $ instead. However, for a given mode some of these
variables are out of phase with respect to the others. Using the
eigenfunctions of all variables as recycling perturbations
simultaneously without taking into account the relative phase between
them does not lead to the excitation of a single mode, but to the
excitation of a sum of different modes (similar to choosing a trial
eigenfunction). From the phase information contained in the complex
FFT of the various variables, we determined that, at least for the
modes we are interested in, the quantities $ v_r $ and $ v_\theta $
have the same phase, while the other two are out of phase by
$ \pi / 2 $ with respect to $ v_r $ and $ v_\theta $.

\begin{figure}
  \includegraphics[width=84mm]{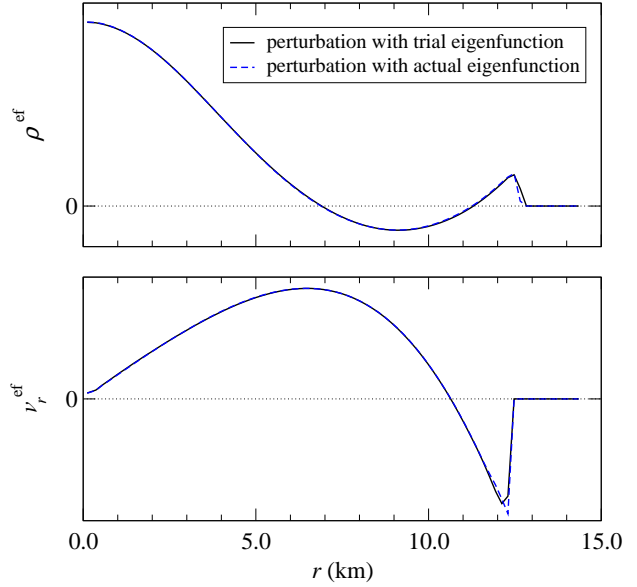}
  \caption{Radial profiles of the eigenfunctions $ \rho^\mathrm{ef} $
    for the $ H_1 $-mode (along the equatorial plane; upper panel) and
    $ v_{\theta}^\mathrm{ef} $ for the $ {}^{2}p_1 $-mode (along
    $ \theta = \pi / 4 $; lower panel) obtained from the evolution of
    model~A1. For an initial perturbation using either a trial
    eigenfunction (black solid lines) or an extracted eigenfunction
    (blue dashed lines), the shapes of the eigenfunctions agree
    well. The eigenfunctions are scaled to the same maximum height.}
  \label{fig:eigenfunction_comparison}
\end{figure}

While the suppression of undesired additional modes in the power
spectrum significantly improves when eigenfunction recycling is
performed, the form of the eigenfunction of various metric and
hydrodynamic quantities extracted from the recycling run is usually
altered only negligibly as compared to the eigenfunction extracted
from the original simulation with the trial perturbation. In
Fig.~\ref{fig:eigenfunction_comparison} we present radial profiles
of the rest-mass density eigenfunction $ \rho^\mathrm{ef} $
along the equatorial plane (upper panel) and of the
$ \theta $-velocity eigenfunction $ v_{\theta}^\mathrm{ef} $
along $ \theta = \pi / 4 $ (lower panel). They are extracted from both
the original simulations of model~A1 using $ l = 0 $
(for $ \rho^\mathrm{ef} $) and $ l = 2 $ (for
$ v_{\theta}^\mathrm{ef} $) perturbations with trial
eigenfunctions, and also from the respective $ H_1 $ and $ {}^{2}p_1 $
recycling runs. Only at the outer stellar boundary the shape of the
eigenfunctions depends slightly on whether the model is perturbed by a
trial or extracted eigenfunction, while in the bulk of the star the
eigenfunctions are practically identical. We find similar results for
the fundamental modes $ F $ and $ {}^{2\!}f $ and for the eigenfunctions
of other metric and hydrodynamic quantities in model~A1 and several
other moderately rotating models.

We can thus conclude that for such rotating stellar models an initial
perturbation with trial eigenfunctions is adequate to precisely
obtain the mode frequency and to extract an accurate corresponding
eigenfunction of the fundamental modes and their first overtones from
the evolution. Additionally, for such models a \emph{single} recycling
run suffices to efficiently suppress the excitation of all unwanted
modes. However, if the peak of the investigated mode in the power
spectrum is small and/or if several modes interact (see also the
discussion of avoided crossings in
Section~\ref{subsection:avoided_crossings}), which is typically the
case for higher order modes in rapidly rotating models, another
recycling loop may be necessary to clearly determine the mode
frequency and eigenfunctions, and to channel most of the initial
perturbation energy into a single oscillation mode.

Particularly in the case of the avoided crossing of the $ H_1 $ and
the $ {}^{4}p_1 $-mode in sequence~A and~B (see also
Figs.~\ref{fig:frequencies_a_au} and~\ref{fig:frequencies_b_bu}), we
use a modified recycling strategy for an accurate mode
analysis. Starting from a model where the mode frequency and
eigenfunctions can still be clearly determined, the next model in the
sequence is perturbed with the eigenfunctions of the investigated mode
extracted from the previous model. This {\em sequential} recycling is
a very helpful tool to resolve problems with ambiguous or unclear mode
frequencies and character of the eigenfunction.


\section{Non-linear pulsations}
\label{section:nonlinear_pulsations}

Although linear perturbations of rotating stars are assumed to have a
vanishingly small amplitude, so that the background equilibrium star
is unaffected by a linear oscillation mode, in certain situations
non-linear effects can become important. Our non-linear evolution code
allows us to investigate such effects, of which we present several
cases in the following.

\begin{figure}
  \includegraphics[width=84mm]{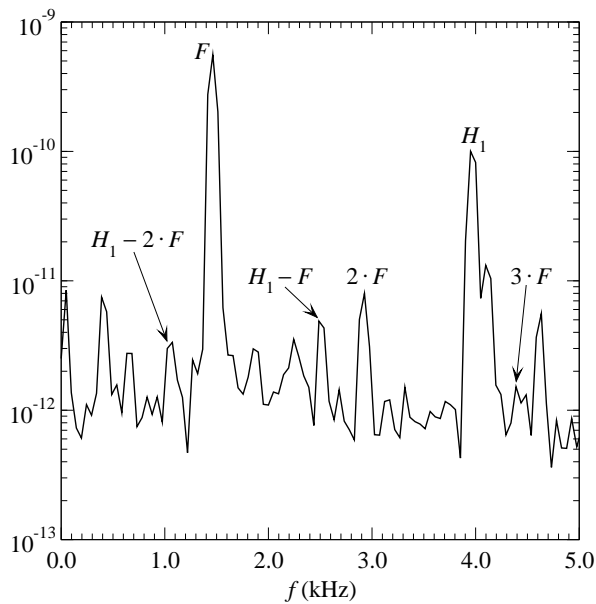}
  \caption{Fourier transform of the evolution of the radial profile of
    the rest-mass density $ \rho $ along the equatorial plane for the
    nonrotating model (A0/AU0/B0/BU0) with an $ l = 0 $ trial
    eigenfunction used as initial perturbation. Additional to linear
    modes, several non-linear harmonics can be identified.}
  \label{fig:nonlinear_harmonics_l=0}
\end{figure}


\subsection{Non-linear harmonics}
\label{subsection:nonlinear_harmonics}

The most basic non-linear effect we see in our simulations is the
appearance of non-linear harmonics of the linear pulsation modes, a
general property of non-linear systems
\citep[cf.][\S~28]{landau_76}. To lowest order, these arise as linear
sums and differences of different linear modes, including
self-couplings. If the system has eigenfrequencies $ \omega_i $, the
non-linearity of the equations excites modes at frequencies
$ \omega_i \pm \omega_j $, with amplitudes proportional to the product
of the amplitudes of the combining frequencies. We note that such
non-linear harmonics were also recently noticed by~\citet{zanotti_05_a}
in the numerical investigation of the dynamics of oscillating,
relativistic, high-density tori around Kerr black holes. 

In Fig.~\ref{fig:nonlinear_harmonics_l=0} we present the Fourier PSD
of the density evolution of the nonrotating model (A0/AU0/B0/BU0), to
which a finite radial $ l = 0 $ initial perturbation of the
form~(\ref{eq:perturbation_l=0}) was added. In addition to the main
linear modes $ F $ and $ H_1 $, one can observe several of their
non-linear harmonics, such as the self-couplings $ 2 \cdot F $,
$ 3 \cdot F $ and the linear sums $ H_1 - F $ and
$ H_1 - 2 \cdot F $. In the PSD a large number of additional peaks can
be seen, and essentially all of those should correspond to non-linear
harmonics of the excited linear modes. It is interesting to note that
several non-linear harmonics are present that have frequencies much
smaller than the fundamental radial mode $ F $ (which, in the linear
approximation, possesses the lowest frequency). These can actually
fall into the frequency range of the inertial modes for rotating
models. Thus, in rotating models further non-linear interactions
between radial modes and inertial modes can be expected (see also the
discussion in Section~\ref{subsection:nonlinear_couplings}).

\begin{figure}
  \includegraphics[width=84mm]{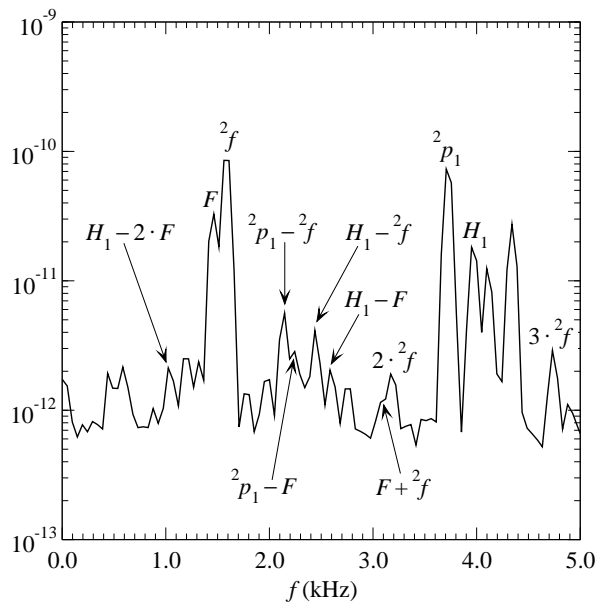}
  \caption{Same as Fig.~\ref{fig:nonlinear_harmonics_l=0}, but with an
    $ l = 2 $ trial eigenfunction used as initial perturbation.}
  \label{fig:nonlinear_harmonics_l=2}
\end{figure}

Non-linear harmonics can occur not only due to couplings between modes
of the same index $ l $ (such as the $ l = 0 $ modes discussed above),
but also due to couplings between modes of different
$ l $. Fig.~\ref{fig:nonlinear_harmonics_l=2} shows several identified
harmonics that represent linear sums and differences between linear
modes for the nonrotating model, when an $ l = 2 $ perturbation was
added to the initial data of the nonrotating model. Due to the
approximate nature of the chosen trial eigenfunction, $ l = 0 $ modes
are also excited in addition to the main $ l = 2 $ modes. The presence
of both $ l = 0 $ and $ l = 2 $ modes then leads to the appearance of
several non-linear harmonics, which include cases like
$ {}^{2}p_1 - F $, $ H_1 - {}^{2\!}f $, etc. It is thus clear that even
though in the linear approximation modes of different $ l $ are
orthogonal to each other (in a nonrotating perfect fluid star),
non-linear effects couple all linear modes with different $ l $ that
are present in a non-linear simulation)\footnote{Notice, however, that
  for relativistic stars there exists no proof yet on the 
  completeness of quasi-normal modes.}.

\begin{figure}
  \includegraphics[width=84mm]{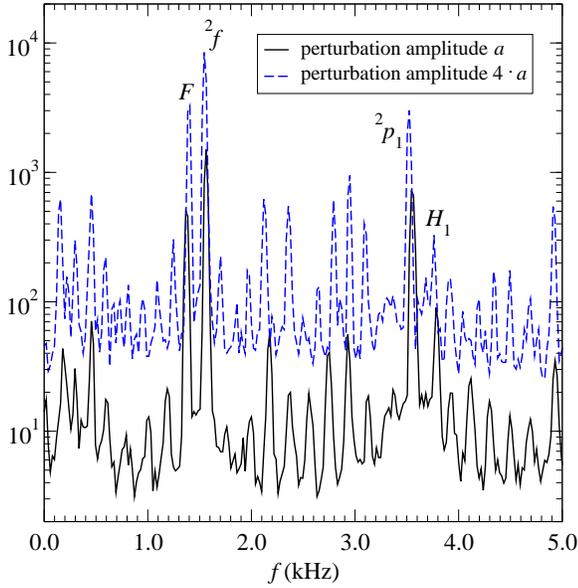}
  \caption{Fourier transform of the evolution of the radial profile of
    the $ \theta $-velocity $ v_{\theta} $ along $ \theta = \pi / 4 $ for
    model~A1 with a combined $ F $, $ H_1 $, $ {}^{2\!}f $, and $
    {}^{2}p_1 $-mode eigenfunction as initial perturbation. Changing
    the perturbation amplitude $ a $ reveals the non-linear scaling of
    the coupled modes as opposed to the linear scaling of the linear
    modes.}
  \label{fig:nonlinear_harmonics_scaling}
\end{figure}

In rotating models, where there can be a perplexing alternation of
various linear modes and non-linear harmonics in a PSD plot, one can
easily distinguish the non-linear harmonics from the linear modes by
comparing PSDs produced from simulations with different initial
perturbation amplitudes. As expected, the linear modes scale almost
linearly, while the non-linear harmonics scale as the product of the
amplitudes of the modes (or of the modes and harmonics) from which
they are produced. Such a case is shown in
Fig.~\ref{fig:nonlinear_harmonics_scaling}, where model~A1 from
sequence~A is used in which the linear modes $ F $, $ H_1 $,
$ {}^{2\!}f $, and $ {}^{2}p_1 $ are all excited at roughly the same
strength using the eigenfunction recycling technique for the initial
perturbation as described in Section~\ref{section:recycling}.
Comparing two simulations of this model that differ by a common factor
of 4 in the initial perturbation amplitude, one can clearly notice
that while the amplitudes of the linear modes scale nearly linearly,
the amplitudes of the various non-linear harmonics scale non-linearly
(with many of them scaling roughly quadratically).

\subsection{Non-linear 3-mode couplings}
\label{subsection:nonlinear_couplings}

\begin{figure}
  \includegraphics[width=84mm]{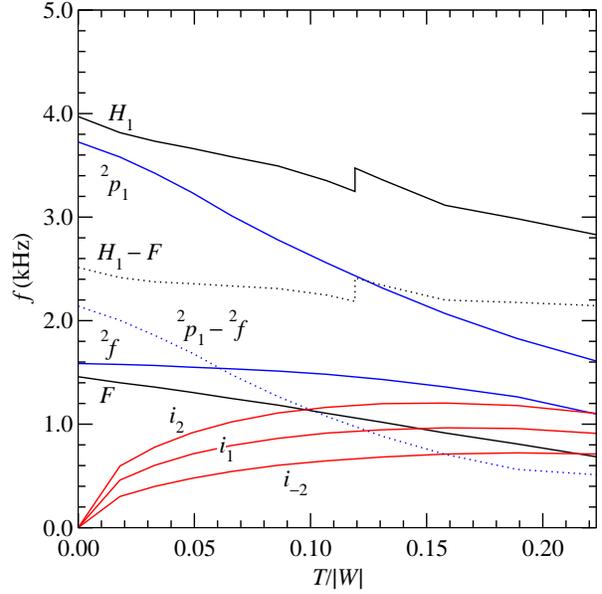}
  \caption{Frequencies of various linear modes and the two non-linear
    harmonics $ H_1 - F $ and $ {}^{2}p_1 - {}^{2\!}f $ for sequence~A.
    At specific rotation rates the frequencies of linear and non-linear
    modes coincide, which can lead to non-linear 3-mode coupling.}
  \label{fig:nonlinear_three_mode_coupling}
\end{figure}

The presence of non-linear harmonics opens the possibility of 3-mode
couplings when the star is rotating. The reason such 3-mode couplings
can take place is the fact that the effect of rotation on the
different modes varies. As was already discussed in
Section~\ref{subsection:rotation_trends}, higher-order modes are
typically affected stronger by rotation than lower-order modes,
which results in avoided crossings between mode sequences. Rotation
also influences the frequencies of the various non-linear harmonics to
a different degree. Thus, at certain rotation rates, a non-linear
harmonic can have the same frequency as a linear mode. Examples
of such cases are demonstrated in
Fig.~\ref{fig:nonlinear_three_mode_coupling}, which shows the
frequencies of several linear modes and of the two non-linear
harmonics $ {}^{2}p_1 - {}^{2\!}f $ and $ H_1 - F $ as a function of
$ T / |W| $. Apparently the $ {}^{2}p_1 - {}^{2\!}f $ harmonic is
crossing the frequency of the fundamental quasi-radial $ F $-mode at
about $ T / |W| \sim 0.1 $, while the $ H_1 - F $ harmonic is crossing
the frequency of the $ {}^{2}p_1 $-mode at about
$ T / |W| \sim 0.12 $. We also note that at any rotation rate one can
expect several non-linear harmonics to coincide in frequency with some
of the infinitely many inertial modes contained in the inertial mode
range.

At such crossings, the coinciding of the frequencies could
potentially lead to resonance effects and even to parametric
instabilities. In such a way, significant energy from one mode
could be transfered to other modes. The most interesting case
would be if pulsational energy from the quasi-radial mode, which
weakly radiates gravitational waves, could be transfered non-linearly
to stronger radiating nonradial modes. Since during core bounce
a significant amount of kinetic energy is stored in the radial
modes of pulsation, the transfer of even a small percentage of
this kinetic energy to a nonradial mode could result in the emission
of strong gravitational waves. This scenario has first been suggested
by \citet{clark_79_a} and \citet{eardley_83_a}, who investigated the
parametric instabilities that could take place, using a Newtonian,
slowly rotating collapse model \citep[for recent related work for
nonrotating or slowly rotating relativistic stars,
see][]{passamonti_05_a}.

In \emph{nonrotating} or \emph{slowly rotating} models, such a
parametric instability can only take place under special conditions
that would allow the two modes to be in resonance. Here we find that
in \emph{rapidly rotating} models rotational shifting of the frequency
of different modes broadens the range of parameters for which
interesting resonances could take place. We particularly notice that
the quasi-radial mode will be in resonance with some inertial mode(s)
for \emph{all rotation rates above a critical value}. It is thus
interesting to further study the possible energy transfer between
different modes excited after, e.g., a core collapse or an
accretion-induced collapse event, either on secular time-scales or as
a parametric instability.

Here we only observe the necessary conditions for non-linear 3-mode
couplings to take place\footnote{We include the case of self
  couplings, when discussing 3-mode couplings.}. Whether such
couplings will indeed lead to strong parametric resonances and
enhanced gravitational wave emission remains to be investigated
through much more detailed studies.


\subsection{Mass-shedding-induced damping}
\label{subsection:mass_shedding}

Another striking example of a non-linear effect is the
mass-shedding-induced damping of oscillations in stellar models which
rotate at or near the mass-shedding limit. This new damping mechanism
was first observed and discussed by SAF, and is especially important
for pulsations in uniformly rotating stars. In SAF, the
mass-shedding-induced damping was demonstrated for the uniformly
rotating models~BU8  and~BU9 using fixed-spacetime evolutions, i.e.\
the Cowling approximation. 

\begin{figure}
  \includegraphics[width=84mm]{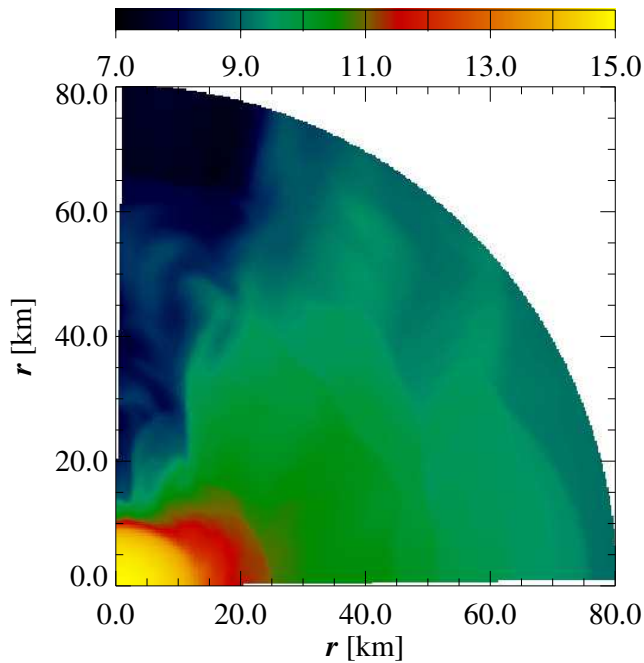}
  \caption{Distribution of the logarithm of the rest-mass density,
    $ \log \rho $ ($ \mathrm{g\ cm}^{-3} $), in the meridional plane
    for model BU9 with ideal fluid EOS at $ t = 4.4 \mathrm{\ ms} $ in
    the Cowling approximation. The ordinate coincides with the
    rotation axis. The single shocks traveling through the atmosphere
    and creating a matter envelope are clearly visible. The black
    color coding corresponds to the low density artificial atmosphere
    where the rest-mass density is actually much smaller than
    $ 10^7 \mathrm{\ g\ cm}^{-3} $. Note that only the innermost 80~km
    of the computational grid are shown.}
  \label{fig:mass_shedding_2d_cowling}
\end{figure}

As explained in SAF, the damping mechanism works as follows. As the
star approaches the mass-shedding limit, the effective gravity near
the equatorial surface diminishes, exactly vanishing at the
mass-shedding limit. A small radial pulsation then suffices to cause
mass-shedding after each oscillation period. As a result, a
low-density envelope is created around the star. This envelope is
initially concentrated in the regions close to the stellar equator,
but with each oscillation period more and more mass is shed in the
form of shock waves, and the envelope expands outwards and away from
the equatorial plane. Since in rotating stars every pulsation mode
also has a radial velocity component, the damping affects all
modes. In SAF, it was found that the damping in the Cowling
approximation can be rather strong. Here we investigate the same
damping effect in the CFC approximation and compare the two cases.

We first repeat the study of the mass-shedding-induced damping for
the radially perturbed model BU9, as presented in SAF, in the Cowling
approximation. Fig.~\ref{fig:mass_shedding_2d_cowling} shows the
distribution of the rest-mass density $ \rho $ in the meridional plane
using the (nonisentropic) ideal fluid EOS~(\ref{eq:ideal_gas_eos}). At
time $ t=4.4\mathrm{\ ms} $ several oscillations have already occurred
and the high-entropy envelope has acquired a near-equilibrium state,
extending into almost the entire computational grid (whose outer
boundary has been set to 5 stellar equatorial radii with 120
additional logarithmically spaced radial grid points)\footnote{Note
  that the equatorial stellar radius is at
  $ r_\mathrm{e} \simeq 17 \mathrm{\ km} $. In
  Table~\ref{table:equilibrium_models} dimensionless units are used
  for $ r_\mathrm{e} $, resulting in a different numerical value of
  the same physical location.}. Near the equatorial plane the envelope
has a rest-mass density of $ \sim 10^{-6}$ to
$ 10^{-4} \rho_\mathrm{c} $, through which further consecutive shocks
propagate. Only at angles $ \lesssim 20^\circ $ (as measured from the
rotation axis) does the high-entropy envelope not completely form.

\begin{figure}
  \includegraphics[width=84mm]{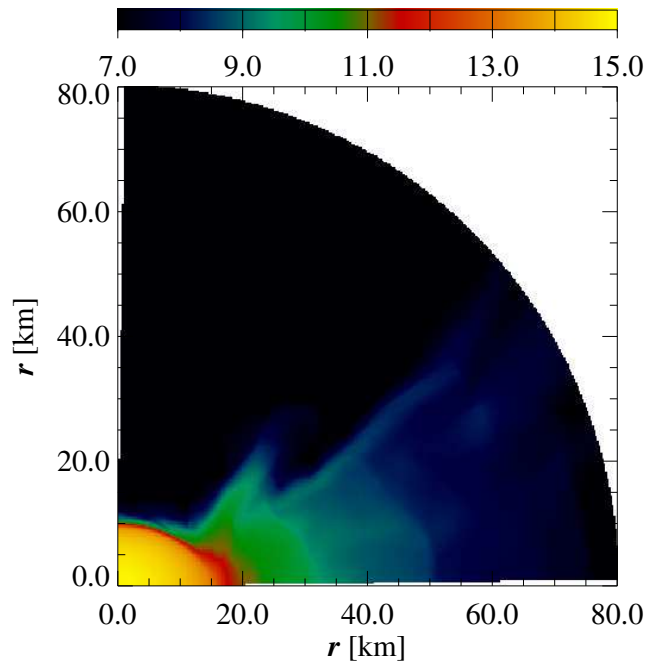}
  \caption{Same as Fig.~\ref{fig:mass_shedding_2d}, but for a
    spacetime coupled to the hydrodynamics (CFC simulation).
    Note that in contrast to the Cowling approximation, a much smaller
    angular part of the atmosphere is filled with a matter envelope,
    which also has significantly also lower density.}
  \label{fig:mass_shedding_2d}
\end{figure}

We then study the same perturbed model (with a similar initial
effective perturbation amplitude) but also evolving the spacetime,
in the CFC approximation. Compared to the Cowling approximation, 
the mass-shedding behavior and the properties of the matter envelope
are now significantly different. In Fig.~\ref{fig:mass_shedding_2d},
one can see (at the same time $ t = 4.4 \mathrm{\ ms} $  as in
Fig.~\ref{fig:mass_shedding_2d_cowling}) that the high-entropy
envelope outside the star is confined to within roughly $ 45^\circ $
with respect to the equatorial plane, filled with matter of one to two
orders of magnitude lower density than in the Cowling approximation.
In addition, the consecutive shocks barely reach the outer grid
boundary.

\begin{figure}
  \includegraphics[width=84mm]{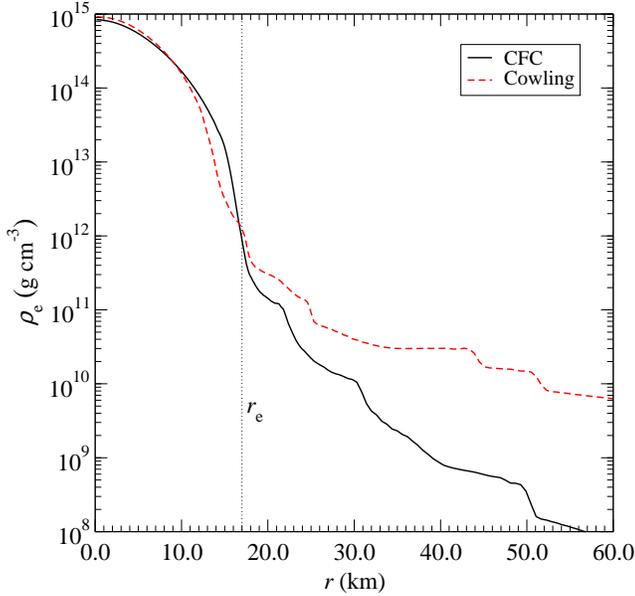}
  \caption{Radial profiles of the rest-mass density
    $ \rho $ along the equatorial plane for model BU9 with
    ideal fluid EOS at $ t = 4.4 \mathrm{\ ms} $ in CFC (black solid
    line) and in the Cowling approximation (red dashed line). The
    single shocks due to mass-shedding are clearly visible. The dotted
    line marks the equatorial stellar radius.}
  \label{fig:mass_shedding_profile}
\end{figure}

Fig.~\ref{fig:mass_shedding_profile} shows the corresponding radial
rest-mass density profiles along the equatorial plane for both the
Cowling and CFC approximations. The envelope in the CFC approximation
has a lower density than in the Cowling approximation, with the
difference consistently increasing with distance and becoming 2 orders
of magnitude at the outer grid boundary. In this plot the single
shocks generated by the stellar oscillations, which first create the
envelope and later propagate through it, are clearly noticeable. For
model~BU9 the frequency of the $ F $-mode oscillations, which are
mainly responsible for the mass-shedding in the above simulations is
about twice as high in the Cowling than in the CFC approximation
($ 2.313 \mathrm{\ kHz} $ vs.\ $ 1.169 \mathrm{\ kHz} $). This is,
however, not reflected in the shock pattern of the radial density
profile in Fig.~\ref{fig:mass_shedding_profile}, where naively twice
as many shocks between $ r_\mathrm{e} $ and some radius
$ r > r_\mathrm{e} $ could be expected. This is a consequence of the
unequal profiles of the shock propagation velocity in the envelope due
to the different density profiles, as well as interactions of shocks
traveling at different velocities due to unequal shock strengths. The
latter effect can also lead to reverse shocks, which further
complicate the picture. Note that in the density profile from the
simulation in the Cowling approximation in
Fig.~\ref{fig:mass_shedding_profile}, the significant loss of the
matter in the stellar regions close to $ r_\mathrm{e} $ by
mass-shedding is evident, whereas in the CFC simulation the initial
density profile inside the stellar boundary is nearly preserved.

\begin{figure}
  \includegraphics[width=84mm]{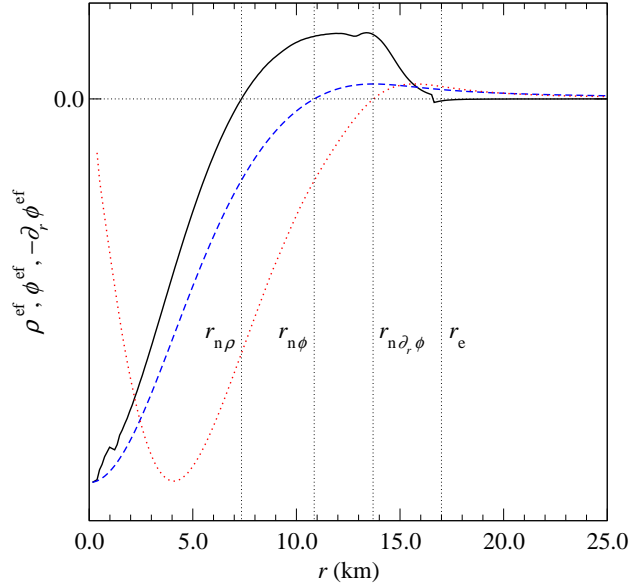}
  \caption{Radial profiles of the $ F $-mode eigenfunction of the
    rest-mass density $ \rho^\mathrm{ef} $ (black solid line), the
    conformal factor $ \phi^\mathrm{ef} $ (blue dashed line), and its
    negative radial derivative $ - \partial_r\phi^\mathrm{ef} $ (red
    dotted line) along the equatorial plane for model BU9 with ideal
    fluid EOS in CFC. The vertical dotted lines mark the locations
    $ r_\mathrm{n\,\rho} $, $ r_\mathrm{n\,\phi} $,
    $ r_{\mathrm{n\,}\partial_r \phi} $ of the eigenfunction nodes and
    the equatorial stellar radius $ r_\mathrm{e} $. The vertical
    scaling of the eigenfunctions is arbitrary, but their sign is
    not.}
  \label{fig:mass_shedding_eigenfunctions}
\end{figure}

In an equilibrium model rotating at the mass-shedding limit, effective
gravity at the equatorial surface vanishes due to a delicate balance
between the pressure gradient, the gravitational force, and the
centrifugal force. In the Cowling approximation the spacetime is held
fixed during the entire simulation, and thus the gravitational force
cannot react to local over- or underdensity caused by the pulsations.
Evidently, in this case matter can easily be ejected from the stellar
equatorial surface at each pulsation. In contrast to this, when the
spacetime is coupled to matter it responds to changes in the
hydrodynamic evolution. The most strongly excited pulsation mode in
the mass-shedding simulations of model BU9 is the $ F $-mode, whose
rest-mass density eigenfunction $ \rho^\mathrm{ef} $ features a node
at $ r_\mathrm{n\,\rho} \simeq 7 \mathrm{\ km} $ in the equatorial
plane\footnote{The occurence of at least one node in the radial
  eigenfunction $ \rho^\mathrm{ef} $ of a quasi-radial mode is a
  consequence of mass conservation.}, located at less than half the
stellar equatorial radius (see
Fig.~\ref{fig:mass_shedding_eigenfunctions}). However, not everywhere
in the star an over- or underdensity is reflected by an according
localised increase or decrease of gravity, as the \emph{local}
gravitational pull is created by the \emph{global} density
distribution. Consequently, the nodes of the eigenfunctions of both
the conformal factor $ \phi^\mathrm{ef} $ and its negative radial
derivative $ - \partial_r \phi^\mathrm{ef} $ (representing the
perturbation of the gravitational force) lie much farther out at
$ r_\mathrm{n\,\partial_r \phi} \simeq 14 \mathrm{\ km} >
r_\mathrm{n\,\phi} \simeq 11 \mathrm{\ km} > r_\mathrm{n\,\rho} $. In
a region $ r_\mathrm{n\,\rho} < r \lesssim r_\mathrm{n\,\phi} $ the
perturbations of density ($ \rho^\mathrm{ef} $) and gravitational
force ($ \phi^\mathrm{ef} $, $ - \partial_r \phi^\mathrm{ef} $)
have opposite sign, and thus at some point during a pulsation cycle
the gravitational force there decreases although the local density
increases. Nevertheless, close to the stellar equatorial surface
both $ \phi^\mathrm{ef} $ and $ - \partial_r \phi^\mathrm{ef} $ are
again in phase with $ \rho^\mathrm{ef} $, and remain greater than zero
(albeit rather small) \emph{at} the surface. Thus, whenever the local
density there increases in a pulsation, this small local net gain
in the gravitational pull suffices to efficiently attenuate
mass-shedding, as observed in the CFC simulations.

\begin{figure}
  \includegraphics[width=84mm]{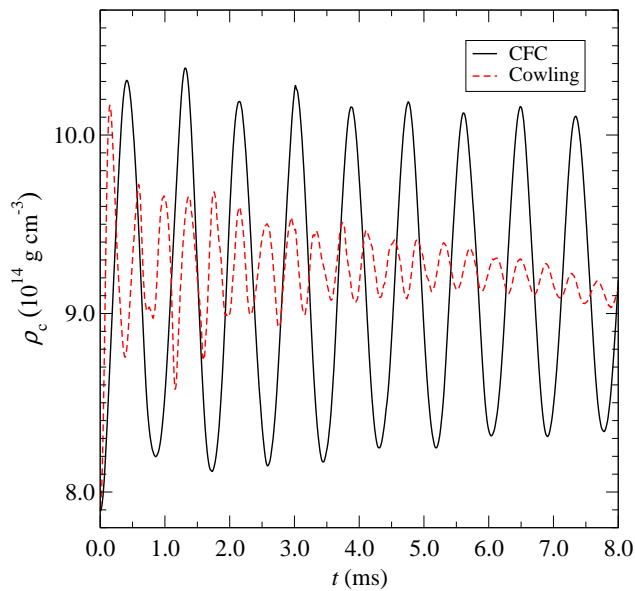}
  \caption{Time evolution of the central rest-mass density
    $ \rho_\mathrm{c} $ for model BU9 with ideal fluid EOS at
    $ t = 4.4 \mathrm{\ ms} $ in CFC (black solid line) and in the
    Cowling approximation (red dashed line). In the Cowling
    approximation, significant mass-shedding-induced damping can be
    seen, whereas the decline of the oscillation amplitude in CFC can
    be mostly attributed to mode-beating.}
  \label{fig:mass_shedding_evolution}
\end{figure}

When matter is shed from the star, kinetic energy is carried away to
the expense of the pulsational energy. Consequently, the pulsations
of the star are gradually damped, depending on the intensity of
mass-shedding. Fig.~\ref{fig:mass_shedding_evolution} shows the
time evolution of the central rest-mass density for model~BU9 obtained
from the CFC simulation and the one in the Cowling approximation. It
is evident that the damping of the pulsations is very strong in the
latter case, leading to an approximate damping time-scale of
$ 10 \mathrm{\ ms} $. In contrast, the insignificant mass-shedding in
the CFC simulation results in only small damping on a much longer
time-scale. Actually, a long-term observation of the time evolution of
$ \rho_\mathrm{c} $ reveals that most of the decline of the
oscillation amplitude in the time window of
Fig.~\ref{fig:mass_shedding_evolution} can be explained by a
mode-beating effect, which superimposes the genuine damping. Even with
high initial perturbation amplitudes we are not able to observe
significant unambiguous mass-shedding-induced damping during typical
evolution times of $ 20 \mathrm{\ ms} $. Simulations with lower
initial perturbation amplitudes or of models which rotate not so close
to the mass-shedding limit exhibit less mass-shedding compared to the
above model, which is in accordance with the findings presented by
SAF.

Even though in the CFC simulation the mass-shedding-induced damping of
pulsations is not as strong as in the Cowling approximation, it could
still have significant implications for unstable modes that grow on
secular time-scales due to the gravitational-radiation driven CFS
instability. In particular, the $ l = m = 2 $ $ f $-mode becomes
unstable only near the mass-shedding limit in uniformly rotating stars
\citep[see][]{stergioulas_98_a, morsink_99_a}. A detailed
investigation is required to determine whether the rate of
mass-shedding-induced damping is shorter than the secular growth rate
of the instability at amplitudes smaller than of order unity.

In Section~\ref{subsection:perturbations} we have demonstrated that
for exciting $ F $-mode oscillations, an initial velocity perturbation
of the form~(\ref{eq:perturbation_l=0}) is more appropriate than the
density perturbation of Eq.~(\ref{eq:perturbation_rho}) if the
spacetime is coupled to the hydrodynamics. However, we have performed
mass-shedding test simulations of model~BU9 in the Cowling
approximation, which show that in this case a velocity perturbation
with typical amplitudes either results in a strong negative or
positive drift in the time evolution of the central rest-mass density,
depending on the sign of the initial perturbation amplitude. Avoiding
this by significantly reducing the initial perturbation amplitude in
turn effectively suppresses the mass-shedding and the resulting
damping of the stellar pulsations. Following SAF, we therefore use the
density perturbation~(\ref{eq:perturbation_rho}) for the simulations
of the mass-shedding model~BU9 in both CFC and the Cowling
approximation presented here. To compensate for the amplification of
the oscillation amplitude during \emph{evolution} using equal
\emph{initial} perturbation amplitudes for a coupled spacetime as
compared to the Cowling approximation (see
Section~\ref{subsection:perturbations}), we use different initial
perturbation amplitudes in CFC ($ a = 0.01 $) than in the Cowling
approximation ($ a = 0.05 $). This yields an approximately equal
height of the first oscillation peak during the evolution (see
Fig.~\ref{fig:mass_shedding_evolution}), so that we can consider the
initial conditions in the two simulations to be similar.


\section{Gravitational waves}
\label{section:gravitational_waves}

In the numerical simulations of the models presented above, we extract
the gravitational waves emitted by the pulsations initiated
in each model. For this we use the Newtonian quadrupole formula in
its time-integrated form \citep[as described in detail
in][]{dimmelmeier_02_b}, which yields the quadrupole wave amplitude
$ A^\mathrm{E2}_{20} $ as the lowest order term in a multipole
expansion of the radiation field into pure-spin tensor harmonics
\citep{thorne_80_a}.


\subsection{Gravitational wave power spectrum}
\label{subsection:wave_spectrum}

\begin{figure}
  \includegraphics[width=84mm]{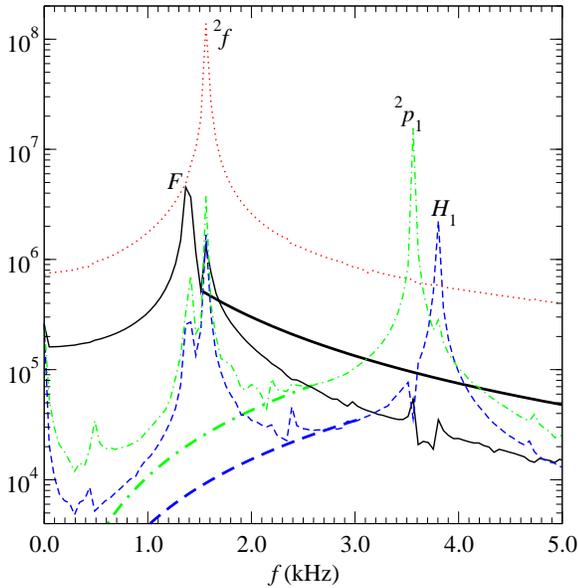}
  \caption{Gravitational wave power spectrum for model~A1 excited with
    recycled eigenfunctions of the $ F $, $ H_1 $, $ {}^{2\!}f $, and
    $ {}^{2}p_1 $-mode, respectively. In the $ F $, $ H_1 $, and
    $ {}^{2}p_1 $-mode recycling runs a significant
    $ {}^{2\!}f $-mode component is also present in the gravitational
    wave signal. The thick lines indicate an artifical fall-off with
    quadratic frequency dependence assumed to get rid of these
    spurious contributions. Units on the $y$-axis are arbitrary.}
  \label{fig:gw_power_spectrum}
\end{figure}

Fig.~\ref{fig:gw_power_spectrum} shows the
gravitational wave power spectrum (i.e.\ the PSD of the time evolution
of $ A^\mathrm{E2}_{20} $) for a series of simulations of model~A1
(i.e.\ the most slowly rotating model of sequence~A) in which the
$ F $, $ H_1 $, $ {}^{2\!}f $, and $ {}^{2}p_1 $-modes are individually
excited using the eigenfunction recycling technique. As already stated
in Section~\ref{section:recycling} the initial excitation amplitude
$ a $ for each individual mode was chosen such that the PSDs of the
integrated density variations in the equatorial plane reach similar
amplitudes (see also lower panel of Fig.~\ref{fig:recycling}).
However, this choice for $ a $ results in a different relative
oscillation amplitude $ \delta \rho_\mathrm{c} $ of the central
density during the evolution (with $ 2 \delta \rho_\mathrm{c} $ being
the density variation measured top to bottom of the oscillation) for
each of the four simulations. In Fig.~\ref{fig:gw_power_spectrum} we
thus scale the original gravitational wave signal such that it
corresponds to $ \delta \rho_\mathrm{c} = 1 \mathrm{\ per\ cent} $.
This is possible as we find that the maximum amplitude
$ |A^\mathrm{E2}_{20}|_\mathrm{max} $ of the quadrupole signal is
approximately proportional to $ \delta \rho_\mathrm{c} $ (and in turn
also to $ a $) to fair accuracy for the four linear stellar pulsation
modes investigated here.

As expected, the quadrupolar $ {}^{2\!}f $-mode is the strongest
emitter of gravitational waves, while the other three modes are still
within roughly one or two orders of magnitude of this dominant
gravitational wave emitting mode. Note that an excitation of the
$ {}^{2\!}f $-mode results in a nearly monochromatic gravitational wave
power spectrum, indicating that the recycled eigenfunctions used for
exciting this mode are of high accuracy. The PSD for the individual
excitations of the $ F $, $ H_1 $, and $ {}^{2}p_1 $-mode are also
dominated by the appropriate frequency of the respective mode, but
exhibit also additional contributions of other excited modes. These
contributions are usually small (at least one order of magnitude
smaller than the dominant mode), but as the quadrupole
$ {}^{2\!}f $-mode strongly radiates gravitational waves, a small
leaking of energy into that mode from the main excitation mode is
sufficient to exhibit a significant contribution at that specific
frequency. This is particularly apparent in the case of the
$ H_1 $-mode, which radiates gravitational waves only through its
\emph{rotationally-induced} $ l \geq 2 $ terms.

In order to suppress the unwanted $ {}^{2\!}f $-mode contribution in
the computation of the characteristic signal frequency and amplitude
in the following Section~\ref{subsection:wave_detectability}, for the 
gravitational wave power spectrum of the $ F $, $ H_1 $, and
$ {}^{2}p_1 $-mode simulations we have assumed a fall-off behavior
with quadratic frequency dependence in the region of the spectrum
where the spurious $ {}^{2\!}f $-mode is located (indicated by thick
lines in Fig.~\ref{fig:gw_power_spectrum}).

We also note that for the excitations of the overtones $ H_1 $ and
$ {}^{2}p_1 $ with recycled eigenfunction, some inertial modes
(visible at the low-frequency end in the PSD of
Fig.~\ref{fig:gw_power_spectrum}) also weakly contribute to the
emission of gravitational waves.


\subsection{Detectability of gravitational waves}
\label{subsection:wave_detectability}

In order to assess the prospects for detectability of the
gravitational wave signal from our models of pulsating neutron stars
by current and planned interferometer detectors, we calculate the
characteristic dimensionless amplitude $ h_\mathrm{c} $ of the signal
from the quadrupole wave amplitude $ A^\mathrm{E2}_{20} $ as
described in \citet{zanotti_05_a}. We perform a Fourier transform of
the transverse traceless gravitational wave signal,
\begin{equation}
  \tilde{h} (f) =
  \int_{-\infty}^\infty \!\! e^{2 \pi i f t} h^\mathrm{TT} (t) \, dt,
  \label{eq:waveform_fourier_tranform}
\end{equation}
where
\begin{equation}
  h^\mathrm{TT} =
  \frac{1}{8} \sqrt{\frac{15}{\pi}} \frac{A^\mathrm{E2}_{20}}{r_\mathrm{gw}}
  \label{eq:amplitude_to_signal_strain}
\end{equation}
in the equatorial plane of the neutron star (assuming optimal
detection geometry of the interferometer) with $ r_\mathrm{gw} $ being
the distance from the emitting source to the detector. To obtain the
detector dependent characteristic signal frequency
\begin{equation}
  f_\mathrm{c} =
  \left( \int_0^\infty \!
  \frac{\langle |\tilde{h}|^2 \rangle}{S_h}
  f \, df \right)
  \left( \int_0^\infty \!
  \frac{\langle |\tilde{h}|^2 \rangle}{S_h}
  df \right)^{-1}
  \label{eq:characteristic_frequency}
\end{equation}
and characteristic signal amplitude
\begin{equation}
  h_\mathrm{c} =
  \left( 3 \int_0^\infty \!
  \frac{S_{h\mathrm{\,c}}}{S_h} \langle |\tilde{h}|^2 \rangle
  f \, df \right)^{1/2}\!\!\!\!\!\!\!\!,
  \label{eq:characteristic_amplitude}
\end{equation}
the power spectral density $ S_h $ of the detector is needed (with
$ S_{h\mathrm{\,c}} = S_h (f_\mathrm{c}) $). We approximate the
average
$ \langle |\tilde{h}|^2 \rangle $ over randomly distributed angles
by $ |\tilde{h}|^2 $. From Eqs.~(\ref{eq:characteristic_frequency},
\ref{eq:characteristic_amplitude}) the signal-to-noise ratio can be
computed as $ S / N = h_\mathrm{c} / [h_\mathrm{rms} (f_\mathrm{c})]$,
where $ h_\mathrm{rms} (f_\mathrm{c}) = \sqrt{f_\mathrm{c} S_h (f_\mathrm{c})} $
is the value of the rms strain noise (i.e.\ the theoretical
sensitivity window) for the detector at the characteristic frequency.

In Table~\ref{table:gw_sensitivity} we give the values of
$ f_\mathrm{c} $ and $ h_\mathrm{c} $ for the gravitational wave
signal emitted by the \emph{most slowly rotating} model~A1 excited
with recycled eigenfunctions, computed for the theoretical power
spectral density $ S_h $ of the detectors VIRGO, LIGO~I and advanced
LIGO located at a distance of $ r_\mathrm{gw} = 10 \mathrm{\ kpc} $
from the source. The strong emission of the gravitational waves by the
quadrupolar $ {}^{2\!}f $-mode and its first overtone $ {}^{2}p_1 $ is
reflected by the large wave amplitudes $ h_\mathrm{c} $ of
$ \sim 400 \times 10^{-22} $ and $ \sim 80 \times 10^{-22} $,
respectively, compared to only $ \sim 20 \times 10^{-22} $ for the
$ F $-mode and $ \sim 10 \times 10^{-22} $ for the $ H_1 $-mode.

\begin{table}
  \centering
  \caption{Characteristic frequency $ f_\mathrm{c} $ and
    characteristic amplitude $ h_\mathrm{c} $ for a gravitational wave
    signal emitted by the \emph{most slowly rotating} model~A1 excited
    with recycled eigenfunctions of the $ F $, $ H_1 $, $ {}^{2\!}f $,
    and $ {}^{2}p_1 $-mode, respectively, as prospectively seen by
    various interferometer detectors. For comparison, the frequency
    $ f $ of the respective stellar oscillation mode is also
    given. All frequencies are in kHz, and $ h_\mathrm{c} $ is given
    in units of $ 10^{-22} $ for a source at
    $ r_\mathrm{gw} = 10 \mathrm{\ kpc} $ oscillating with
    $ \delta \rho_\mathrm{c} = 1 \mathrm{\ per\ cent} $. The signal
    duration time is assumed to be
    $ t_\mathrm{gw} = 20 \mathrm{\ ms} $.}
  \label{table:gw_sensitivity}
  \begin{tabular}{@{}l@{\q}r@{\qq}r@{\q}r@{\qq}r@{\q}r@{\qq}r@{\q}r@{}}
    \hline
    & &
    \multicolumn{2}{c@{\qq}}{VIRGO} &
    \multicolumn{2}{c@{\qq}}{LIGO~I} &
    \multicolumn{2}{c@{}}{adv.~LIGO} \\
    Mode &
    \multicolumn{1}{c@{\qq}}{$ f $} &
    \multicolumn{1}{c@{\q}}{$ f_\mathrm{c} $} &
    \multicolumn{1}{c@{\qq}}{$ h_\mathrm{c} $} &
    \multicolumn{1}{c@{\q}}{$ f_\mathrm{c} $} &
    \multicolumn{1}{c@{\qq}}{$ h_\mathrm{c} $} &
    \multicolumn{1}{c@{\q}}{$ f_\mathrm{c} $} &
    \multicolumn{1}{c@{}}{$ h_\mathrm{c} $} \\
    \hline
    $ F $         & 1.400 & 1.370 &  19.3 & 1.338 &  18.9 & 1.236 &  17.4 \\
    $ H_1 $       & 3.816 & 3.783 &  12.1 & 3.792 &  12.1 & 3.791 &  12.1 \\
    $ {}^{2\!}f $ & 1.577 & 1.524 & 463.3 & 1.467 & 467.5 & 1.292 & 393.8 \\
    $ {}^{2}p_1 $ & 3.580 & 3.545 &  76.9 & 3.545 &  76.9 & 3.544 &  76.9 \\
    \hline
  \end{tabular}
\end{table}

Note that we integrate $ h^\mathrm{TT} $ over a finite duration
time of $ t_\mathrm{gw} = 20 \mathrm{\ ms} $ in the FFT and scale
$ \delta \rho_\mathrm{c} $ to 1 per cent for each individual excited
mode. To compute $ h_\mathrm{c} $ in a straightforward way from the
values given in Table~\ref{table:gw_sensitivity} for an arbitrary
source distance, signal duration, and central oscillation amplitude of
the star, the following simple scaling laws can be used:
$ h_\mathrm{c} \propto r_\mathrm{gw}^{-1} $ and
$ h_\mathrm{c} \propto \sqrt{t_\mathrm{gw}} $, while
$ h_\mathrm{c} \propto |A^\mathrm{E2}_{20}|_\mathrm{max} $ for a given
power spectral density $ S_h $ and thus approximately proportional
to $ \delta \rho_\mathrm{c} $ (see discussion in
Section~\ref{subsection:wave_spectrum}).

As we artificially remove the spurious contribution of the quadrupole
$ {}^{2\!}f $-mode to the gravitational wave power spectrum of the
$ F $, $ H_1 $, and $ {}^{2}p_1 $-mode recycling runs (see
Section~\ref{subsection:wave_spectrum}), the small discrepancy between
the values for characteristic gravitational wave signal frequency
$ f_\mathrm{c} $ and the stellar oscillation mode frequency $ f $ in
Table~\ref{table:gw_sensitivity} is predominantly caused by the
dependence of $ f_\mathrm{c} $ on the power spectral density
$ S_h $ of the specific detector. Accordingly, as
$ h_\mathrm{c} \neq h^\mathrm{TT} $, the values of characteristic
amplitudes $ h_\mathrm{c} $ are slightly different for each detector.
The effect of $ S_h $ on $ f_\mathrm{c} $ and $ h_\mathrm{c} $ is
strongest for the gravitational wave signal emitted by the relatively
low frequency modes $ F $ and $ {}^{2\!}f $.

\begin{figure}
  \includegraphics[width=84mm]{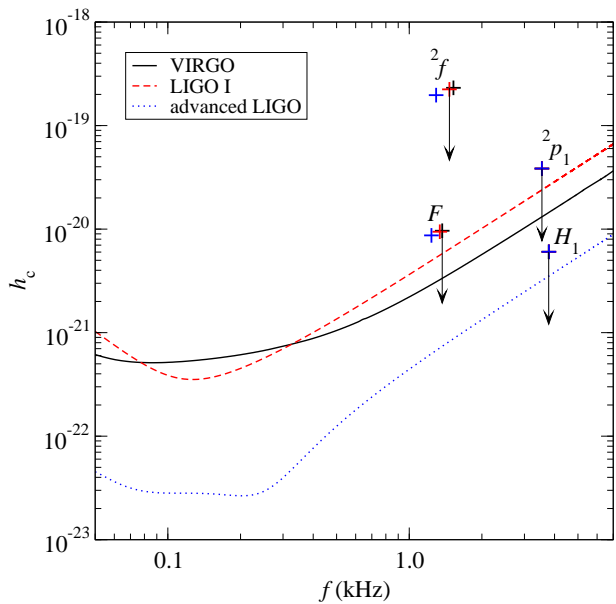}
  \caption{Distribution of the characteristic gravitational wave
    signal amplitude $ h_\mathrm{c} $ in the frequency space for
    model~A1 excited with recycled eigenfunctions of the $ F $,
    $ H_1 $, $ {}^{2\!}f $, and $ {}^{2}p_1 $-mode, respectively.
    The crosses represent
    $ h_\mathrm{c} $  for a source at
    $ r_\mathrm{gw} = 10 \mathrm{\ kpc} $, oscillating with
    $ \delta \rho_\mathrm{c} = 5 \mathrm{\ per\ cent} $. The signal
    duration time is assumed to be
    $ t_\mathrm{gw} = 20 \mathrm{\ ms} $. The  sensitivities
    $ h_\mathrm{rms} $ of the interferometer detectors VIRGO (black
    solid line), LIGO~I (red dashed line), and advanced LIGO (blue
    dotted line) are also shown. Particularly for the $ F $ and
    $ {}^{2\!}f $-mode signal, the value of $ h_\mathrm{c} $ and
    $ f_\mathrm{c} $ is slightly dependent on the sensitivity of the
    respective interferometer detector, as indicated by the color
    coding. The arrows indicate the location of $ h_\mathrm{c} $ for $
    \delta \rho_\mathrm{c} = 1 \mathrm{\ per\ cent} $.}
  \label{fig:gw_sensitivity}
\end{figure}

In Fig.~\ref{fig:gw_sensitivity} we show the location of the
characteristic gravitational wave amplitude $ h_\mathrm{c} $ for the
$ F $, $ H_1 $, $ {}^{2\!}f $, and $ {}^{2}p_1 $-mode signal of model~A1
in relation to the sensitivity windows of VIRGO, LIGO~I, and advanced
LIGO. We again assume a distance to the source of
$ r_\mathrm{gw} = 10 \mathrm{\ kpc} $, i.e.\ within our Galaxy, and
(in contrast to Table~\ref{table:gw_sensitivity}) an oscillation
amplitude $ \delta \rho_\mathrm{c} = 5 \mathrm{\ per\ cent} $. In such
a scenario the gravitational wave signal from a neutron star pulsating
in the $ {}^{2\!}f $-mode with the adopted oscillation amplitude would
be clearly detectable even by current interferometer observatories
with a signal-to-noise ratio of around 50 for VIRGO and LIGO~I (and
300 for the planned advanced LIGO detector). On the other hand, the
$ F $ and $ {}^{2}p_1 $-mode signals lie only marginally above the
sensitivity threshold of current detectors, corresponding to a
signal-to-noise ratio of at most $ \sim 2 $. The wave signal from the
$ H_1 $-mode, in turn, could possibly be detectable only by the
proposed advanced LIGO detector for the selected pulsation amplitude
and duration.

If $ \delta \rho_\mathrm{c} $ is reduced to 1 per cent, as exemplified
by the arrows in Fig.~\ref{fig:gw_sensitivity}, the $ F $ and
$ {}^{2}p_1 $-mode signals clearly fall out of the sensitivity window
of VIRGO and LIGO~I, but would be detectable by advanced LIGO, while
the $ H_1 $-mode signal is now evidently out of the range even for
the latter. Only the signal from the $ {}^{2\!}f $-mode still exhibits a
signal-to-noise ratio of around 10 for the current detectors. It is
important to note that using the scaling properties of the signal
discussed above, the data presented in Fig.~\ref{fig:gw_sensitivity}
can be used to determine the necessary minimum stellar oscillation
amplitude $ \delta \rho_\mathrm{c} $, maximum source distance
$ r_\mathrm{gw} $, and minimum signal duration time $ t_\mathrm{gw} $
required for a prospective detection (set by a specific
signal-to-noise ratio above 1) of the gravitational wave emitted by
model~A1.

We also emphasize that the frequencies of the investigated linear
pulsation modes $ F $, $ H_1 $, $ {}^{2\!}f $, and $ {}^{2}p_1 $ are
located at the rapidly ascending high frequency slope of the detector
sensitivity curves. Consequently, for the fixed rest mass sequences~A
and~AU the decrease of the frequencies for these modes with increasing
rotation can enhance the prospects for detection by shifting the
signals towards the more sensitive domain of the detector sensitivity
window. For the fixed central density sequences~B and~BU, however, the
opposite dependence of the $ H_1 $ and particularly of the
$ {}^{2\!}f $-mode frequencies on rotation reverses this effect for these
two modes.

Further increase of the signal-to-noise ratio of the gravitational
wave signals presented here could be achieved with the advanced LIGO
detector, using its planned ability to be tuned at higher frequencies
in a narrow-banding mode. The chances for a successful detection of
neutron star oscillation modes will also rise dramatically by the
simultaneous operation of high-frequency-band detectors such as the
proposed dual sphere detectors \citep{cerdonio_01_a}, which are
designed to have an improved sensitivity at frequencies of a few kHz
and would thus be particularly suitable for detecting the first
overtones $ H_1 $ and $ {}^{2}p_1 $.


\subsection{Gravitational wave asteroseismology}
\label{subsection:gw_asteroseismology}

It is clear from the discussion of our results so far, that pulsating
rotating neutron stars are gravitational wave sources that depend on
several parameters (EOS, mass, angular momentum, differential rotation
law, initial perturbation amplitude, damping mechanisms, etc.). All
these parameters may have different effects on the oscillation
spectrum of the star and, therefore, the successful extraction of the
physical characteristics of the source from the gravitational wave
signal will be difficult to achieve. It is important to isolate each
effect on the gravitational waveform in order to find general trends.

Empirical formulas that can be used for gravitational wave
asteroseismology (with the aim of pin-pointing the correct
high-density EOS through gravitational wave observations) were
constructed in recent years \citep{andersson_98_a, kokkotas_01_a, 
benhar_04_a}, assuming nonrotating cold equilibrium models. Thus,
these formulas contain only the effect of EOS and mass. More recently,
finite temperature effects and slow-rotation effects on the mode
frequencies were computed \citep{ferrari_03_a, ferrari_03_b,
  ferrari_04_a}, but these effects have not yet been incorporated in
the empirical formulas for asteroseismology. In our work we focus on
the effects of rotation, using a particular cold EOS as an example. A
complete treatment will, of course, require using a large sample of
realistic EOSs, but such an attempt is beyond the scope of this
work. Using only the frequencies extracted from the simulations
presented here, we can nevertheless identify several interesting
features that indicate qualitatively the influence of rotation on the
empirical formulas used in gravitational wave asteroseismology.

First, the frequencies of most modes show a simple dependence on
rotation so that the empirical formulas obtained for nonrotating stars
could be extended to rotation by multiplying with a factor containing
only one or two powers of a rotational parameter such as
$ T / |W| $. The exception is the $ H_1 $-mode, which will require
special treatment due to the avoided crossing at moderate values of
$ T / |W| $. Second, the frequencies of most modes, such as the $ F $,
$ {}^{2\!}f $ and $ H_1 $-mode, become nearly independent of the degree
of differential rotation when parametrized as a function of
$ T / |W| $. However, the frequency of the $ {}^{2}p_1 $-mode is
strongly affected by the degree of differential rotation. Thus, an
additional parameter, measuring the strength of differential rotation,
could be built into an empirical formula for this specific mode.

Ideally, if the four modes $ F $, $ H_1 $, $ {}^{2\!}f $, and
$ {}^{2}p_1 $ were detected, then a set of empirical formulas for
their frequencies, constructed in the way described above, would allow
the simultaneous extraction of the mass, radius, $ T / |W| $, and
degree of differential rotation. Of course, this ideal situation may
become more complicated by the inclusion of finite temperature
effects, which would add at least one more parameter to the system.


\section{Summary and Outlook}
\label{section:summary}

Using the axisymmetric general relativistic hydrodynamics code
\textsc{CoCoNuT} we have investigated pulsations of uniformly and
differentially rotating neutron star models. We have compared our
numerical simulations, in which the spacetime dynamics is coupled to
the evolution of the fluid, to previous results performed under the
assumption of a fixed spacetime (Cowling approximation). In the 
present work we have used the so-called conformal flatness condition
(CFC) for the spatial three-metric, this being an excellent and
well-tested approximation of the exact spacetime in the regime studied
here. The coupled system of (hyperbolic) hydrodynamics and (elliptic)
metric equations has been solved using the novel approach of combining
Riemann-solver-based HSRC methods for the fluid evolution and spectral
methods for the computation of the spacetime
metric~\citep{dimmelmeier_05_a}.

As equilibrium initial data we have constructed four sequences of
relativistic polytropes with parametrized rotation, which have then
been perturbed by small amplitude $ l = 0 $, 2, or 4 trial
eigenfunctions to excite pulsation modes. By analyzing the time
evolutions of various hydrodynamic and metric quantities using Fourier
transforms along radial profiles or on the entire spatial grid, we
have obtained the pulsation frequencies for the fundamental
quasi-radial ($ l = 0 $) $ F $-mode, the fundamental quadrupole
($ l = 2 $) $ {}^{2\!}f $-mode, and their respective overtones, the
$ H_1 $ and $ {}^{2}p_1 $-mode, as well as three inertial modes
labelled $ i_{-2} $, $ i_1 $, and $ i_2$. Additionally, for two
differentially rotating sequences we have obtained the frequencies for
the $ {}^{4}f $ and $ {}^{4}p_1 $-mode. We have found that for these
two sequences the $ {}^{4}p_1 $-mode engages in an avoided crossing
with the $ H_1 $-mode. This is a consequence of the different
influence of rotation on the frequency of these two modes, which
brings their frequencies closer with increasing rotation rate. At the
avoided crossing, an exchange of the character of eigenfunctions of
the $ H_1 $ and $ {}^{4}p_1 $-mode has been observed. Although linear
perturbation theory predicts the existence of an infinite number of
inertial modes in a finite frequency range, in our simulations we have
detected the predominant excitation of only a few specific inertial
modes, which are excited as by-products of the excitation of other
modes. For the differentially rotating sequence of fixed rest mass,
the inertial mode frequencies reach a maximum value before the
mass-shedding limit is reached and then decrease again.

In order to suppress the simultaneous excitation of more than one
mode, which is typically the case when trial eigenfunctions are used
as initial perturbations, we have employed a new technique, which we
call \emph{eigenfunction recycling}. For this, we first extract the
two-dimensional eigenfunction of velocity components at the frequency
of the mode under consideration and then apply it as an initial
perturbation in a second simulation of the original neutron star
model. This selective excitation of modes works very well even for
overtones.

When comparing the frequency curves of the $ F $-mode for all four
investigated sequences with previous results in the Cowling
approximation, we have observed that the latter typically overestimate
the correct frequency by about a factor of 2 (corresponding to an
absolute difference of $ \sim 1 \mathrm{\ kHz} $), which is consistent
with similar findings for nonrotating stars by
\citet{yoshida_97_a}. Moreover, we have found no evidence for the
$ F $-mode splitting found by SAF to occur in the Cowling
approximation. Thus, we conclude that this effect is an artifact which
can be attributed to the Cowling approximation, as was suggested in
SAF. On the other hand, for the $ H $, $ {}^{2\!}f $, and
$ {}^{2}p_1 $-mode frequencies, much closer agreement with the Cowling
results of SAF has been found. For these modes the relative difference
is usually less than $ \sim 20 \mathrm{\ per\ cent} $. Concerning
empirical formulas constructed by \citet{font_02_a} and SAF using
results in the Cowling approximation in order to predict mode
frequencies for rotating sequencies, we have found that the changing
characteristics of equilibrium models along various sequences
significantly affect the relative differences between the actual
frequencies and those obtained in the Cowling approximation. As a
result, such empirical relations can only be of limited validity,
while still representing a significant improvement compared to the
Cowling approximation.

Even though we have used only small-amplitude initial perturbations,
the use of a non-linear evolution code has allowed us to investigate
various non-linear effects in pulsating neutron star models. We have
observed non-linear harmonics of the linear pulsation modes of our
stellar models, arising as linear sums and differences of various
linear modes, including self-couplings \citep[for similar findings in
high-density accretion tori around Kerr black holes,
see][]{zanotti_05_a}. For rotating models such non-linear harmonics
actually fall into the frequency range of the inertial modes and could
interact with linear modes in non-linear 3-mode couplings in the form
of a resonance or a parametric instability. It would be interesting to
investigate under which conditions strong mode couplings or
instabilities can occur.

In SAF a new non-linear damping mechanism for non-linear pulsations
was found to operate in uniformly rotating models that are near the
mass shedding limit. The damping of pulsations is due to mass
shedding, when fluid elements near the equator become unbound due to
the radial component of the velocity perturbation. This
mass-shedding-induced damping could have severe consequences for,
e.g., the $ f $-mode gravitational-wave driven CFS instability, which
grows on a secular time-scale. Here we have found that while the actual
mass-shedding-induced damping does not occur on dynamical time-scales
(as in the Cowling approximation) it is still present on secular
time-scales, so that a detailed comparison to the $ f $-mode growth
rate is required in order to determine the outcome of the $ f $-mode
instability.

Taking into account that in the core collapse simulations by
\citet{dimmelmeier_01_a, dimmelmeier_02_a, dimmelmeier_02_b} the
dominant quadrupole contribution had an effective gravitational wave
amplitude that was roughly one to two orders of magnitude above the
sensitivity curve of the advanced LIGO detector, it is likely that in
a favourable detection event several pulsation modes of the
proto-neutron star may be detected simultaneously. We have thus
investigated the gravitational wave signals (in the Newtonian
quadrupole approximation) emitted by the linear modes $ F $, $ H_1 $,
$ {}^{2\!}f $, and $ {}^{2}p_1 $, for a slowly rotating neutron star
model. In order to estimate the prospects for detectability of the
signal by interferometric detectors, we have computed the (weakly
detector dependent) characteristic signal frequency and amplitude for
the theoretical power spectral density of VIRGO, LIGO~I, and advanced
LIGO. We infer that for a pulsation amplitude of the stellar central
density of 5 per cent a gravitational wave signal from the
investigated modes (with the exception of the $ H_1 $-mode) is
measurable by current detectors, if the signal is integrated over a
time of at least $ 20 \mathrm{\ ms} $ and the source is located in our
Galaxy. We have shown how scaling properties of the signal can be used
to determine the necessary minimum stellar pulsation amplitude,
maximum source distance, and minimum signal duration time required for
a prospective detection. We have also discussed how effects of
rotation can qualitatively influence empirical formulas used in
gravitational wave asteroseismology and can thus help to constrain
parameters of neutron star models in the case of a successful
detection of gravitational waves from pulsation modes. A more complete
study of the detectability of various modes for all equilibrium models
of our sequences will appear elsewhere.

In the relativistic core collapse simulations by
\citet{dimmelmeier_01_a, dimmelmeier_02_a, dimmelmeier_02_b} the
quasi-periodic gravitational waves emitted around core bounce were
found to have frequencies less than roughly $ 1.1 \mathrm{\ kHz} $. On
the other hand, for nonrotating models constructed with a large sample
of different realistic EOSs, the frequencies of the fundamental
$ l = 2 $ quadrupole $ f $-mode range from
$ \sim 1.35 \mathrm{\ kHz} $ for extremely stiff EOSs to
$ \sim 3.6 \mathrm{\ kHz} $ for extremely soft EOSs \citep[see,
e.g.,][]{andersson_98_a}. One of the main effects lowering the
frequency of emitted gravitational waves is rotation. In our
differentially rotating sequence of fixed rest mass, the nonrotating
model has a fundamental $ l = 2 $ $ f $-mode frequency of
$ \sim 1.6 \mathrm{\ kHz} $, which decreases to
$ \sim 1.1 \mathrm{\ kHz} $ at the largest rotation rate. Therefore,
additional effects must be taken into account in order to explain the
wider range of frequencies found by \citet{dimmelmeier_01_a,
  dimmelmeier_02_b}. One such important factor is the high entropy in
a proto-neutron star immediately after core bounce. The proto-neutron
star is surrounded by a hot envelope into which the pulsations of the
interior can penetrate, increasing the period of each pulsation
\citep[see][]{ferrari_03_a}. Thus, a natural extension of our
investigation will be to compute the frequencies of the various
axisymmetric modes for a realistic proto-neutron star structure.


\section*{Acknowledgments}

It is a pleasure to thank Theocharis Apostolatos, John Friedman,
Kostas Kokkotas, Ewald M\"uller, Christian Ott, Jos\'e Pons, and
Luciano Rezzolla for helpful comments and discussions. We are grateful
to Olindo Zanotti for providing us with data for producing the
sensitivity curves of the interferometer detectors. Financial support
for this research has been provided by the EU ILIAS initiative of the
European Network of Theoretical Astroparticle Physics (ENTApP), the
German Research Foundation DFG (SFB/Transregio~7
`Gravitations\-wellen\-astro\-nomie'), the Greece--Spain bilateral
research grant by the General Secretariat for Research and Technology
(GSRT) and the Spanish Ministerio de Educaci\'on y Ciencia (grants
AYA2004-08067-C03-01 and HG2004-0015), the Pythagoras,
Pythagoras-II, and Heraclitus grants of the Greek Ministry of Education
and Religious Affairs, and the IKYDA German--Greek research
travel grant by DAAD and IKY. The computations were performed at the
Max-Planck-Institut f\"ur Astrophysik in Garching and the
Max-Planck-Institut f\"ur Gravitationsphysik in Golm, Germany.


\label{lastpage}

\end{document}